\documentclass[12pt,a4paper,final]{iopart}

\usepackage{iopams}
\usepackage{graphicx}
\usepackage[breaklinks=true,colorlinks=true,linkcolor=blue,urlcolor=blue,citecolor=blue]{hyperref}
\usepackage{braket}

\begin{document}

\title{Perspectives of quantum annealing: Methods and implementations}

\author{Philipp Hauke$^{1,2}$, Helmut G.~Katzgraber$^{3,4,5}$, Wolfgang Lechner$^{6,7}$\footnote{Corresponding author: Wolfgang.Lechner@uibk.ac.at}, Hidetoshi Nishimori$^{8,9}$, William D. Oliver$^{10,11}$}


\address{$^1$ Kirchhoff-Institute  for  Physics,  Heidelberg  University,  69120  Heidelberg, Germany 
}
\address{$^2$ Institute  for  Theoretical  Physics,  Heidelberg  University,  69120  Heidelberg,  Germany
}
\address{$^3$Microsoft Quantum, Microsoft Corporation, Redmond, WA
98052, USA}
\address{$^4$Department of Physics and Astronomy, Texas A\&M University,
College Station, Texas 77843-4242, USA}
\address{$^5$Santa Fe Institute, 1399 Hyde Park Road, Santa Fe, New Mexico 87501 USA}
\address{$^6$ Institute for Theoretical Physics,  University of Innsbruck,
6020 Innsbruck, Austria }
\address{$^7$ Institute for Quantum Optics and Quantum Information, Austrian Academy of Sciences, 6020 Innsbruck, Austria
}
\address{$^8$Institute of Innovative Research, Tokyo Institute of Technology, Nagatsuta-cho, Midori-ku, Yokohama 226-8503, Japan}
\address{$^9$Graduate School of Information Sciences, Tohoku University, Sendai 980-8579, Japan}

\address{$^{10}$Research Laboratory of Electronics, Massachusetts Institute of Technology, Cambridge, Massachusetts 02139, USA}
\address{$^{11}$MIT Lincoln Laboratory, 244 Wood Street, Lexington, Massachusetts 02420, USA}

\begin{abstract}
Quantum annealing is a computing paradigm that has the ambitious goal of efficiently solving large-scale combinatorial optimization problems of practical importance. However, many challenges have yet to be overcome before this goal can be reached. This perspectives article first gives a brief introduction to the concept of quantum annealing, and then highlights new pathways that may clear the way towards feasible and large scale quantum annealing. Moreover, since this field of research is to a strong degree driven by a synergy between experiment and theory, we discuss both in this work. An important focus in this article is on future perspectives, which complements other review articles, and which we hope will motivate further research. 
\end{abstract}

\vspace{2pc}

\section{Introduction}

The last two decades have seen tantalizing  progress in the engineering of quantum devices, raising the hope to realize quantum technologies based on precise control over large ensembles of microscopic quantum degrees of freedom \cite{kielpinski2002architecture,blais2004cavity,wallraff2004strong,schoelkopf2008wiring,clarke2008superconducting,bakr2009quantum,Ladd2010,o2010quantum,chow2010detecting,blatt2012quantum,cirac2012goals,Yan2016,bernien2017probing}. 
Notable examples are small-scale prototypes of circuit(gate)-based quantum computers, which use logical gate operations on quantum bits (qubits). 
These devices, if ideally built to a large scale, are theoretically proven to be able to run certain quantum algorithms exponentially faster than any classical computer running classical algorithms. However, qubit-based quantum computers are extremely hard to scale up in practice \cite{preskill2018quantum}, with current quantum processing units consisting of less than one hundred qubits. Even worse, the aforementioned statement assumes that these are perfect qubits; including error correction results in a large overhead when encoding logical variables in physical ones. Given these challenges, researchers have thus searched for less demanding alternatives, which may enable solving certain problems of practical importance, hopefully efficiently given particular criteria. Over the history of classical computing, analog special-purpose machines may be seen to have heralded programmable digital universal silicon-based computers.
In this article, we present perspectives for an important example of a similar nature, but in the quantum world---quantum annealing and adiabatic quantum optimization---in which remarkable experimental and theoretical advances are currently being observed \footnote{We will also draw connections to another such example, namely special-purpose quantum computers known as quantum simulators, which may help researchers to solve quantum many-body problems \cite{blatt2012quantum,cirac2012goals,Hauke2011d,Georgescu2014}.}. 
The potential strength of such an approach---at least in the near to medium term---may be seen in the example that the largest number ever factorized on a quantum computer was done using an adiabatic protocol \cite{Peng2019}. 

\begin{figure} %
\centering
\includegraphics[width=1.0\textwidth]{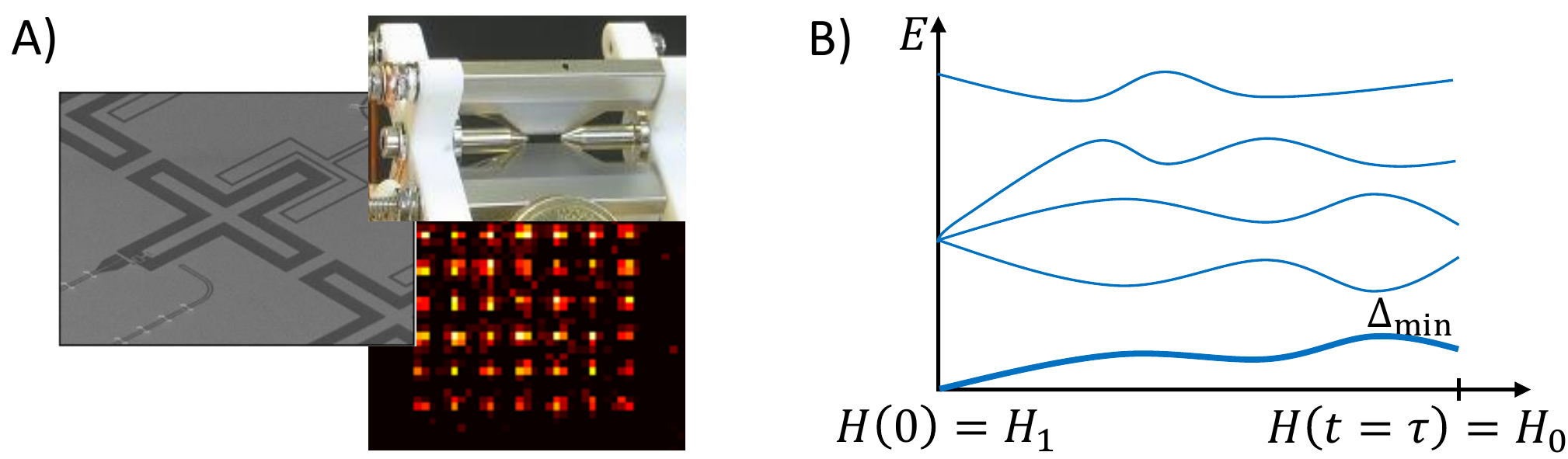}
\caption{(A) Recent years have seen exciting developments on quantum-technology platforms (exemplified here by superconducting qubits, trapped ions, Rydberg atoms). These enable the implementation of quantum annealing protocols with the aim of solving hard optimization problems. (B) Sketch of time-dependent energy spectrum. The solution to the optimization problem is encoded in the ground state of a problem Hamiltonian $H_0$. It is reached at the end time $\tau$ of a slow sweep starting from the ground state of a Hamiltonian $H_1$ that is simple to prepare. If the sweep is sufficiently adiabatic (i.e., slow as compared to an inverse polynomial of the minimum gap $\Delta_\mathrm{min}$), the system remains in the instantaneous eigenstate throughout (thick line). This article discusses experimental as well as theoretical prospects to boost the performance of such quantum annealers. 
Picture credits panel A (clockwise from left): MIT Lincoln Laboratory; Blatt group, University of Innsbruck; LCF, Institut d’Optique, CNRS. 
} 
\label{fig:concept of review}
\end{figure}

Quantum annealing \cite{Nishimori1998,farhi:00,kadowaki:98a} has been designed to solve classical combinatorial optimization problems, with applications ranging from computer science problems \cite{farhi2001quantum}, classification \cite{neven2009training}, quantum chemistry \cite{babbush2014adiabatic}, machine learning \cite{lloyd2013quantum}, search engine ranking \cite{garnerone2012adiabatic} to protein folding \cite{perdomo2012finding}. 
Such optimization problems require the minimization of a cost function, a task that can be rephrased as finding the ground state of a classical Ising Hamiltonian $H_0$ \cite{lucas:14}. 
Many problems of practical importance, however, have cost functions with a large number of local minima, corresponding to Ising Hamiltonians that are reminiscent of classical spin glasses \cite{binder:86,young:98,nishimori:01,mezard:87}. 
These characteristics make it extremely difficult for classical algorithms to find the global minimum \cite{kadowaki:98a}. 
Quantum annealing was conceived as an alternative to solve this formidable task, based on the idea to elevate the classical Ising Hamiltonian $H_0$ to the quantum domain, i.e., by taking it to describe a collection of interacting qubits. 
According to the adiabatic theorem of quantum mechanics \cite{messiah1961quantum,Amin2009} (see also Refs.~\cite{Jansen2007,Lidar2009,Cheung2011}), the ground state of the classical Ising model can be found by initializing the system in the ground state of some initial Hamiltonian $H_1$, which is easy to prepare both theoretically and experimentally. $H_1$ is chosen such that it  does not commute with $H_0$, and the system parameters are changed sufficiently slowly such that the Hamiltonian changes gradually from $H_1$ to $H_0$. 
More explicitly, the system is described by a time-dependent Hamiltonian (see Fig.~\ref{fig:concept of review}) 
\begin{equation}
    \label{eq:H(t)}
    H(t) = A(t) H_0 + B(t) H_1\,,
\end{equation} 
where $B(t)$ is slowly reduced from the initial value $B(0)=1$ to the final value $B(\tau)=0$, with $\tau$ being the computation time,  while $A(t)$ is slowly increased from $A(0)=0$ to $A(\tau)=1$, thus changing the Hamiltonian from $H(0)=H_1$ to $H(\tau)=H_0$. 
The adiabatic theorem states that, under a sufficiently slow change of the parameters in the Hamiltonian, the system ends up in a state close enough to the ground state of the final Hamiltonian $H_0$ if the initial condition is given as the ground state of the initial Hamiltonian $H_1$, meaning that the desired solution to the optimization problem is obtained. Since the final Hamiltonian $H_0$ is a classical Ising model with only commuting operators $\{\sigma_i^z\}$, the $z$ component of the Pauli matrix, the solution can then be read out as the state of the individual qubits in the computational basis.

Thanks to the simplicity and elegance of this approach as well as its potential impact, several organizations including private companies are heavily investing efforts and resources toward its realization as large-scale quantum devices, mainly based on superconducting qubits \cite{bunyk:14,johnson:11,dickson:13,Weber2017,Rosenberg2017}. 
Since the first quantum annealing device, the D-Wave device, has become commercially available in 2011  \cite{johnson:11}, a large number of proof-of-principle demonstrations have appeared, see, for example, Refs.~\cite{albash:18,katzgraber:14,ronnow:14,katzgraber:15,hen:15a,king:15,mandra:16b,mandra:17a,mandra:18} and references therein.

Nevertheless, there is an ongoing scientific debate whether such quantum annealing devices can offer an actual speedup over classical computers \cite{Altshuler2010,boixo:14,ronnow2014defining,Katzgraber2018}. For the case of the simple transverse-field Ising model, where we choose $H_1$ as the transverse field Hamiltonian
$    -\sum_{i=1}^N \sigma_i^x$, 
with $\sigma_i^x$ being the Pauli matrix at qubit $i$, and where $H_0$ is written only in terms of $\{\sigma_i^z\}$, an essential speedup over classical algorithms may not be easy to achieve because equilibrium properties of the model, as well as certain aspects of dynamics, can usually be simulated efficiently by quantum Monte Carlo methods \cite{isakov2016understanding}, although with some exceptions, e.g., Ref.~\cite{Andriyash2017}.  This property is shared by a class of models called `stoquastic Hamiltonians' \cite{Bravyi2009}, which are defined as those having a matrix representation with all off-diagonal elements being non-positive in a product basis, usually the computational basis to diagonalize $\{\sigma_i^z\}$, leading to a classical representation without the sign problem for simulations. The situation is further complicated by the presence of noise and a variety of imperfections in real devices, all of which degrade the performance, sometimes significantly.

In the present article, we aim to highlight new pathways and perspectives for the development of quantum annealing in the hope to clear the way towards establishment of feasible approaches for efficiently solving large-scale problems of practical importance. Therefore, an important focus of this article is on future perspectives, making its characteristics rather complementary to typical review articles \cite{RevModPhys.90.015002,das:08,Venegas-Andraca2018}. Correspondingly, the list of references may be far from complete, for which we apologize to many authors.

This article is organized as follows. 
In Sec.~\ref{sec:theoretical_framework}, we lay some theoretical framework of quantum annealing, starting from the representation of combinatorial optimization problems as classical Ising spin glasses, followed by an explanation of the essence of the adiabatic condition and the connection of the minimum energy gap to computational complexity. 
We also discuss how quantum annealing performs as compared to classical algorithms, and we point out the open question regarding the role of quantum entanglement in any possible quantum speedup. 
The core of the article consists of Secs.~\ref{sec:perspectives_methods} and~\ref{sec:implementations}. 
In Sec.~\ref{sec:perspectives_methods}, we highlight several promising routes towards enhancing the performance of quantum annealing based on improvements in algorithms. These include judicious choices of the quantum driving term $H_1$ and non-adiabatic schemes. 
In Sec.~\ref{sec:implementations}, we discuss perspectives on the superconducting qubit platform and illustrate potential alternatives based on ultracold Rydberg atoms and trapped ions. 
Finally, we briefly summarize our discussions in Sec.~\ref{sec:discussion}.  

\section{Theoretical framework}
\label{sec:theoretical_framework}

In this section, we recapitulate some of the theoretical framework that underlies the concept of quantum annealing, which will be useful for subsequent discussions. 

\subsection{From optimization problems to spin glasses}
\label{sec:optimization problems}

As stated in the Introduction, the main motivation for building quantum annealing machines such as the D-Wave device is to solve combinatorial optimization problems, in particular in the form of binary optimization problems.  While the restriction to binary variables might seem narrow at first, a wide variety of NP-hard optimization problems of interest in industry fall under this category. Because in most cases only heuristic approaches to tackle these hard problems are known, there has been much interest in harnessing quantum effects in an attempt to find better solutions. Here, a `better solution' can signify a cost closer to optimality, faster reaching of optimality at fixed cost, or a more diverse solution pool (if the problem has more than one minimizing configuration).
Prominent potential applications range from spin glasses \cite{venturelli:15a}, to lattice protein
models \cite{perdomo:12}, circuit fault diagnosis
\cite{perdomo:15b,perdomo:17y}, planning \cite{rieffel:15}, job-shop
scheduling \cite{venturelli:15b}, machine learning \cite{benedetti:16},
molecular similarity in chemistry \cite{hernandez:17}, or the optimal
trading trajectory problem \cite{rosenberg:16}, to name a few in addition to those listed in the Introduction. These problems can be cast in a form expressed in terms of binary (Boolean) variables $x_i \in \{0,1\}$, generally as high-order polynomial unconstrained binary optimization problems (HOBO) written as $k$-local interactions with $k\ge 3$\footnote{A $k$-local interaction is a term in the cost function, the function to be minimized, proportional to a product of $k$ binary variables, $x_{i_1}x_{i_2}\cdot\dots\cdot x_{i_k}$}. However, due to manufacturing constraints, current experimental devices can only handle $2$-local interactions, i.e., {\em quadratic} unconstrained binary optimization problems (QUBOs) with a cost function of the form
\begin{equation}
H_{\rm QUBO} = \sum_{ij}Q_{ij}x_ix_j + \sum_i c_i x_i\,.
\label{eq:hgk_qubo}
\end{equation}
The problem to be optimized is then fully specified by $Q_{ij}$ and $c_i$. A broad class of paradigmatic optimization problems from vertex covers to the traveling salesperson problem have been mapped to QUBO form. For a comprehensive study, see Ref.~\cite{lucas:14}.

If the problem of interest has a cost function of high-order interactions ($k$-local with $k\ge 3)$, one should reduce it to QUBO format by using ancillary variables to implement it on a real device. For example, a $3$-local expression $x_1x_2x_3$ is reduced to $x_1x_4$ if we define $x_4=x_2x_3$. The latter condition can be imposed by an additional term in the cost function
\begin{equation}
    3x_4+x_2x_3-2x_2x_4-2x_3x_4.
\label{eq:ancillary_cost}
\end{equation}
This expression is $0$ only when $x_4=x_2x_3$ and is positive (higher cost) otherwise.  Other expressions than Eq.~(\ref{eq:ancillary_cost}) are possible for the same purpose with an appropriate coefficient for each term.   While this technique has the advantage that any problem can in principle be studied on a currently-available device, it usually leads to a large overhead in the number of variables. Therefore, effort should further be made in studying combinatorial problems in their native HOBO form. We concentrate on QUBO in this article.

Equation (\ref{eq:hgk_qubo}) can be conveniently mapped to an Ising-like
expression with the transformation $\sigma_i^z = 1-2x_i$:
\begin{equation}
H_{\rm SG} = -\sum_{ij}J_{ij}\sigma_i^z\sigma_j^z - \sum_i h_i \sigma_i^z+{\rm const.},
\label{eq:hgk_sg}
\end{equation}
where
\begin{eqnarray}
J_{ij} &=& -\frac{1}{4}Q_{ij} ,\\
h_i    &=& -\frac{1}{2}\left(c_i + \sum_j Q_{ij}\right) ,
\end{eqnarray} 
and $\sigma_i^z$ is regarded at this stage as a classical variable taking the values
$\pm 1$.
The Hamiltonian $H_{\rm SG} $ is a typical example of the classical Ising Hamiltonian $H_0$ introduced in the Introduction.
In Eq.~(\ref{eq:hgk_sg}) the constant (physically irrelevant) energy shift
\begin{equation}
E_0 = \frac{1}{4}\left( 2\sum_i c_i + \sum_{ij} Q_{ij}\right) 
\end{equation} 
has been neglected. If the couplers $J_{ij}$ are chosen from a random distribution, the Ising model given in Eq.~(\ref{eq:hgk_sg}) is also known as a spin glass. 
In what follows and without loss of generality, quantum annealing is
discussed from the point of view of spin-glass Hamiltonians. 
We note, however, that the approach can be extended to other physical systems. 
There is a solid body of work on the disordered magnetic systems described as spin glasses, see, e.g., Refs.~\cite{binder:86,young:98,nishimori:01,stein:13} and references therein. Yet,
despite decades of study, our theoretical understanding is still insufficient, and simulations are restricted to small system sizes because 
the inherent frustration and disorder result in optimization problems of 
superb difficulty. This challenge has led to the development of sophisticated
algorithms to study the thermodynamic behavior of classical spin glasses and, more
recently, the adaptation of these methods as heuristics to minimize the energy
of these systems \cite{katzgraber:03f,katzgraber:04c,zhu:16y}. We later
discuss the performance of quantum annealing compared to these new, 
`quantum inspired' optimization techniques. 

\begin{figure} %
\centering
\includegraphics[width=1.0\textwidth]{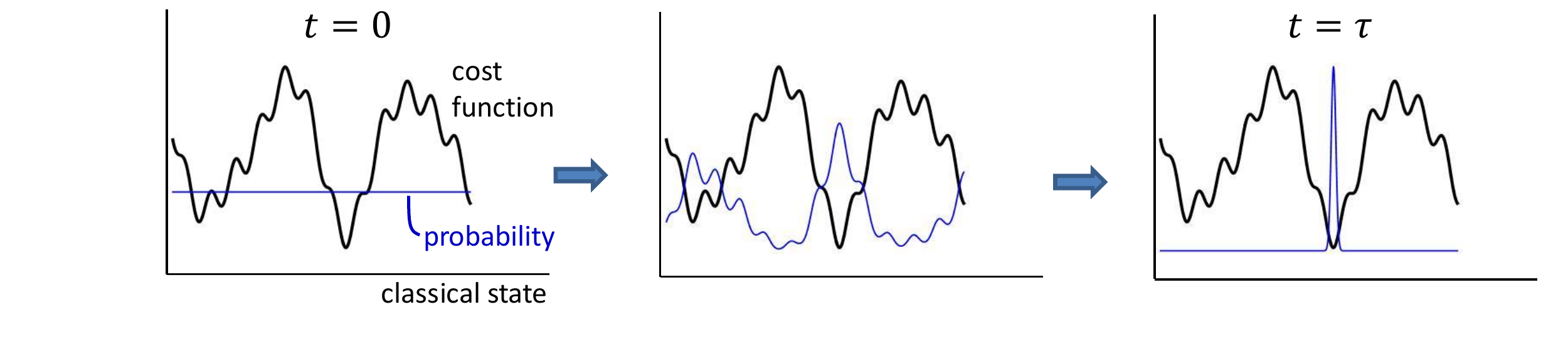}
\caption{
Sketch of quantum annealing protocol for a spin-glass problem. 
A typical spin-glass problem is characterized by a rough cost function (black line) as a function of the classical configurations, i.e., the energy eigenstates of the Ising Hamiltonian $H_{\rm SG}$ (ordered on x-axis).   
While the probability (blue line) to find any given classical state is uniform at the beginning of the annealing protocol, it converges to the desired solution at the end of the sweep, provided that the sweep is sufficiently adiabatic. 
} 
\label{fig:spin glass}
\end{figure}

In quantum annealing, the classical Ising variables $\sigma_i^z$ are promoted to quantum spin operators, i.e., Pauli matrices. The solution to the optimization problem encoded in the ground state of Eq.~(\ref{eq:hgk_sg}) is then sought by slowly sweeping the system from a simple Hamiltonian $H_1$, whose ground state can be easily prepared and which does not commute with $H_{\rm SG}$, to $H_0=H_{\rm SG}$, see Eq.~(\ref{eq:H(t)}). During this procedure, the probability to find a given classical configuration converges from a uniform distribution to a distribution that is ideally strongly peaked at the ground state of $H_{\rm SG}$, see Fig.~\ref{fig:spin glass}. In this context, the difficulty of solving the spin-glass problem translates into the presence of exponentially many energy gaps that are exponentially small once the annealing procedure enters the spin-glass phase \cite{Altshuler2010,Knysh2016}. This makes quantum annealing a formidable task, for theoretical simulation and experimental realization alike. 

\begin{figure} %
\centering
\includegraphics[width=1.0\textwidth]{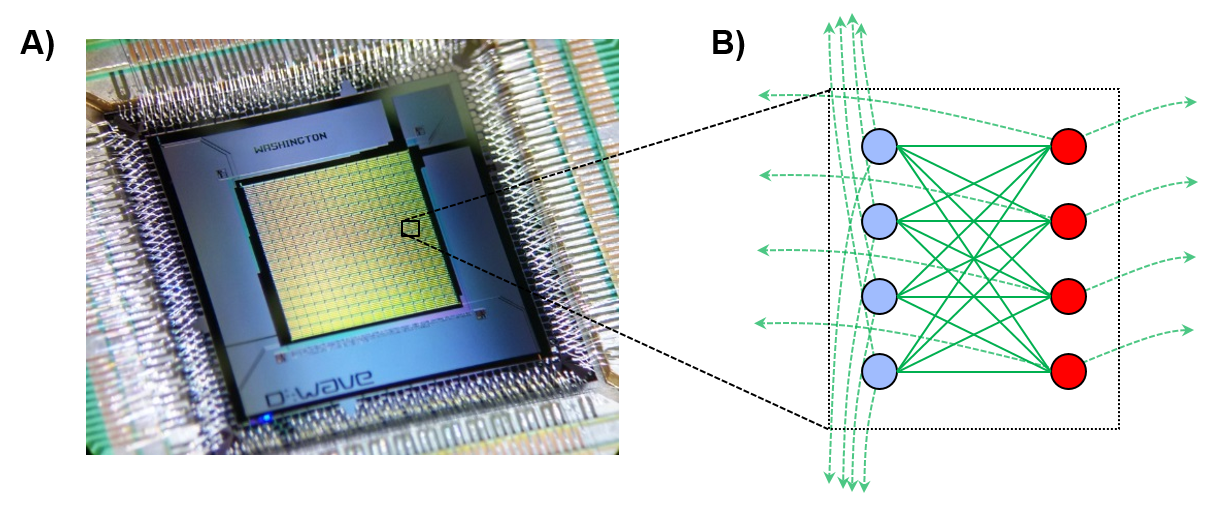}
\caption{D-Wave 2000Q and the Chimera graph. (A) Photograph of a D-Wave 2000Q Washington chip with 2048 qubits. From the web page of D-Wave Systems Inc. \cite{Dwave2000q}. (B) Chimera graph unit cell and connectivity. Bullets indicate qubits within the unit cell and solid lines connections between them. Dashed arrows indicate connections to adjacent unit cells, which are arranged in a square-lattice pattern.} 
\label{fig:Washington}
\end{figure}

Moreover, because the connectivity graph of the currently-available quantum annealers is
limited, the study of any problem that does not happen to match, e.g., the
underlying Chimera graph structure of the D-Wave device (see Fig.~\ref{fig:Washington}) requires `embedding' \cite{choi:08}, i.e., a representation of the logical optimization problem in the physically available graph structure. This embedding typically consumes a considerable overhead in terms of spin variables \cite{Zaribafiyan2017}.
For example, in the extreme case of a fully-connected graph, 
approximately $30$ logical spin variables can be embedded in the $2000$
variables of the D-Wave 2000Q quantum annealer. This overhead severely
limits any asymptotic scaling analysis. As such, random spin-glass
problems defined on the native Chimera lattice of the D-Wave device \cite{bunyk:14} 
have been
extensively used to benchmark these machines, as well as quantum
annealing in general
\cite{katzgraber:14,ronnow:14,katzgraber:15,hen:15a,king:15,mandra:16b,mandra:17a,mandra:18,albash:18}.

An alternative encoding of optimization problems for quantum annealing is the LHZ (Lechner-Hauke-Zoller) scheme \cite{lechner2015quantum,rocchetto2016stabilizers}. In this architecture, the optimization problem is encoded in the local fields acting on the individual qubits, and all interactions are problem independent 4-body terms among nearest neighbors on a two dimensional grid. See Sec. \ref{subsec:non-adiabatic} for more details. Using LHZ, all-to-all models can be implemented in various platforms including transmon qubits \cite{leib2016transmon}, Kerr non-linear oscillators \cite{Goto2015,puri2017quantum}, flux qubits \cite{chancellor2017circuit} and Rydberg atoms \cite{glaetzle2017coherent} which will be discussed in Section \ref{sec:rydberg} in detail. 

\subsection{Adiabatic approximation and adiabatic theorem}
\label{sec:adiabatic_approximation}

As already suggested, the existence of quantum phase transitions and small energy gaps is one of the main sources of difficulties in quantum annealing.  We describe this point in this and the following part of this section.

One of the important theoretical bases of quantum annealing is the adiabatic approximation \cite{messiah1961quantum,Amin2009}. The statement is that a quantum system with slowly time-dependent parameters in the Hamiltonian closely follows the instantaneous eigenstate in its time evolution if we start the dynamics in one of the eigenstates of the initial Hamiltonian.\footnote{We distinguish the term `adiabatic theorem' from `adiabatic approximation', the former being reserved for rigorous mathematical theorems \cite{Jansen2007,Lidar2009,Cheung2011} whereas the latter is for approximations \cite{messiah1961quantum,Amin2009}.} Let us write the instantaneous eigenvalue equation of a Hamiltonian as
\begin{equation}
\label{eq:H(t)parameter}
 H(t)|j(t)\rangle =\epsilon_j(t)|j(t)\rangle,
\end{equation}
where $j=0$ denotes the instantaneous ground state and $j\ge 1$ represents instantaneous excited states. We assume for simplicity that states are discrete and non-degenerate. In Eq.~(\ref{eq:H(t)parameter}), the time variable $t$ is fixed and is regarded as a parameter. 

The adiabatic approximation is concerned with the solution $|\psi(t)\rangle$ to the time-dependent Schr\"odinger equation generated by $H(t)$ with $t$ now a dynamical variable.  
Our interest focuses mostly on the ground state, and thus we assume that the time evolution of the system starts from the ground state of the initial Hamiltonian $H(0)=H_1$, $|\psi(0)\rangle=|0(0)\rangle$. Then, the adiabatic approximation is that the wave function is considered to stay close to the instantaneous ground state after time evolution to $t=\tau$,
$
|\langle 0(\tau)|\psi(\tau)\rangle |^2\approx 1\,,
$
as long as the adiabatic condition holds
\begin{equation}
\max_{0\le t\le\tau} \frac{\left|\langle j(t)|\frac{dH(t)}{dt} |0(t)\rangle \right|}{\Delta_{j,0}(t)^2} \ll 1 \quad (j\ge 1)\,.
\label{AC1}
\end{equation}
Here, $\Delta_{j,0}(t) =\epsilon_j(t)-\epsilon_0(t)$ is the energy gap between the ground state and the $j$th excited state.

The adiabatic condition, Eq.~(\ref{AC1}), is derived from the evaluation of excitation probabilities and is not necessarily mathematically rigorous. Indeed, counterexamples are known  in which the state of the system deviates significantly from the instantaneous ground state even when the adiabatic condition is satisfied.  In particular, when the Hamiltonian has a time-dependent parameter with a time scale different from the intrinsic time scale related to the energy scale of the system, the system sometimes displays an oscillatory behavior even when the adiabatic condition, Eq.~(\ref{AC1}), is satisfied \cite{Amin2009}.

This last possibility is excluded if the time dependence of the Hamiltonian is parametrized by a single variable $s=t/\tau$, where $t$ changes from 0 to $\tau$ and accordingly $s$ runs from 0 to 1, $\tilde{H}(s)\equiv H(t)$.  If the Hamiltonian has this property, which is the case in most theoretical models of quantum annealing, the adiabatic condition Eq.~(\ref{AC1}) for $j=1$ is rewritten as, 
\begin{equation}
 \tau \gg \max_{0\le s\le 1} \frac{\left|\langle 1(s)|\frac{d\tilde{H}(s)}{ds} |0(s)\rangle \right|}{\Delta_{1,0}(s)^2} .
\label{AC2}
\end{equation}
This equation gives an estimate of the computation time $\tau$ in terms of the minimum energy gap $\Delta_{\rm min}=\min_{s}\Delta_{1,0}(s)$ and is frequently used in theoretical analyses.

There exist several mathematically rigorous statements (adiabatic theorems) that give sufficient conditions for the system to stay sufficiently close to the instantaneous ground state \cite{Marzlin2004,Jansen2007,Lidar2009,Cheung2011}. Nevertheless, since our interest is in the qualitative properties of the computation time $\tau$ as a function of the system size $N$, and rigorous adiabatic theorems lead to qualitatively the same conclusions, either polynomial or exponential time dependence of $\tau$, we simply use the adiabatic condition of Eq.~(\ref{AC2}) in the following discussion.

\subsection{Adiabatic quantum computing and quantum annealing}

Often, the framework of quantum computation that imposes the adiabatic condition Eq.~(\ref{AC2}), such that the system is supposed to always follow the instantaneous ground state, is called adiabatic quantum computing. The idea was first proposed in the context of a typical optimization problem of Exact Cover \cite{Farhi2001}, and provides a framework that is known in its most general form to be theoretically equivalent to the traditional circuit(gate)-based model of quantum computing \cite{Aharonov2008,Biamonte2008}. 

Quantum annealing is a related but broader concept \cite{Nishimori1998} \footnote{
Several papers had proposed the term `quantum annealing' \cite{Apollini1988,Apolloni1989,Finnila1994} (see also \cite{Amara1993,Tanaka2000}) before Ref.~\cite{Nishimori1998} formulated quantum annealing in its current style. Those early papers used classical processes following, typically, the imaginary-time Schr\"odinger equation.}. Quantum annealing allows not only strict adiabatic processes but also non-ideal situations including diabatic (non-adiabatic) transitions due to fast changes of Hamiltonian parameters as well as noisy environments of thermal and other origins. Nevertheless, the term quantum annealing is often used in the restricted sense of adiabatic computing, and we follow this tradition as long as it does not lead to a confusion.

\subsection{Energy gap and computational complexity}
\label{subsection:energy_gap}

As seen in Eq.~(\ref{AC2}), performance of quantum annealing strongly depends on the properties of the energy gap between ground and first excited state, $\Delta_{1,0}$, which should not be too small lest the necessary computation time $\tau$ becomes large or the system departs from the instantaneous ground state. 
However, the ground states of the initial and final Hamiltonians are qualitatively quite different because the initial state is supposed to be trivial, while the final state should encode a highly non-trivial optimization problem. This means that a phase transition occurs during the annealing procedure, and a quantum phase transition is known to be characterized by a vanishing energy gap in the thermodynamic limit \cite{sachdev:99}. 

Let us denote the minimum gap $\min_{s}\Delta_{1,0}(s)$ encountered during the annealing protocol as $\Delta_{\mathrm{min}}$. Suppose that this gap closes exponentially as a function of the system size, $\Delta_{\mathrm{min}}\propto e^{-cN}~(c>0)$, which occurs at a first-order quantum phase transition \footnote{A few exceptions are known \cite{Cabrera1987,Laumann2012,Tsuda2013}.}. The order of magnitude of the numerator on the left-hand side of Eq.~(\ref{AC2}) is the same as that of the Hamiltonian, i.e., $\mathcal{O}(N)$, and thus can be ignored in comparison with the much-stronger exponential dependence of the denominator. The computational complexity (the size dependence of the computation time) is then exponential, $\tau \propto e^{2cN}$. In this case, the problem is considered hard to solve.  If, on the other hand, the minimum gap closes polynomially $\Delta_{\mathrm{min}}\propto N^{-l}~(l>0)$, as is usually observed in a second-order phase transition according to finite-size scaling \cite{nishimori:11}, then the computation time is polynomial $\tau \propto N^{2l+1}$ and the problem is considered easy. Therefore the size dependence of the gap and the order of transition play important roles in the analysis of quantum annealing in the adiabatic framework. Notice here that a phase transition is a phenomenon that occurs only in the thermodynamic limit. Nevertheless, the order of transition in this limit gives us important information on the computational complexity for large but finite-size problems\footnote{There are known instances of relevant optimization problems where the minimum gap, e.g., is polynomial for small systems and becomes a stretched exponential only for sufficiently large problem instances \cite{Knysh2016}.}.  
Consequently, one approach to improving the performance of quantum annealers is by mitigating the gap closing, e.g., by judicious choice of the annealing schedule \cite{Zeng2015b} or by inhomogeneous driving of the transverse fields \cite{Susa2018a,Susa2018b}, see Sec.~\ref{sec:Inhomogeneous driving}.

\subsection{Quantum annealing vs classical simulation}

A driving force for the development of quantum annealers is the hope to outperform classical devices, a question that is intrinsically tied to the scaling of the minimum gap. 
Before one can compare quantum annealing to classical approaches, however, one must define the metrics to be used. There are
two main tracks to compare the performance of algorithms: Resource
estimation and performance for specific tasks.

Resource estimation \cite{mcgeoch:12} amounts to determining the
necessary resources a particular method needs to achieve a target value.
These resources can be, for example, time, energy, or memory. The target
value can either be a particular energy threshold or the true optimum of
a benchmark instance. In the case of quantum annealing, it has been
common to use spin-glass-like Hamiltonians as benchmark problems with
the resources measured being the time it takes to solve an ensemble of
problems as a function of the size of the input (here, the number of
variables). Most recently, it has been argued \cite{mandra:18} that
energy should be included in such metrics. In addition, careful
definitions of speedup are needed between quantum and classical
paradigms. The first careful definition of quantum speedup was done in
Ref.~\cite{ronnow:14a}, later extended in Ref.~\cite{mandra:16b} to add
more granularity. 

One can also use specific tasks to evaluate the performance of an
algorithm. For example, one could analyze the quality of the solutions,
as well as how correlated or uncorrelated a set of solutions are when
the algorithms are run multiple times. Multiple studies
\cite{matsuda:09,mandra:17} have shown that quantum annealing is
intrinsically biased and is therefore not a good sampler. As such, we
focus here on resource requirements. However, it is worth emphasizing
that advanced control techniques as outlined in this review, as well as
non-stoquastic drivers, might alleviate the sampling problem for selected problems.

There have been multiple studies on the performance of quantum annealing
and quantum annealing machines
\cite{santoro:02,santoro:06,dickson:13,pudenz:13,smith:13,boixo:13a,ronnow:14,katzgraber:14,lanting:14,santra:14,shin:14,boixo:14,albash:15,albash:15a,katzgraber:15,martin-mayor:15,pudenz:15,hen:15a,venturelli:15a,vinci:15,zhu:16,king:17}.
However, to date, it remains controversial if there is any ``{\em
quantum speedup}'' or not. Such a statement strongly depends on the
benchmark problem used and, to date, there is no industrial application
where quantum annealing outperforms classical heuristics to the best of our knowledge. To extend the
scaling regime of the different benchmark studies, researchers have
shifted their focus to simple basic spin-glass problems, ideally on the native
topology of quantum annealing machines. Unfortunately, it was
demonstrated that random spin-glass benchmarks \cite{ronnow:14} on
sparse graphs might not be complex enough to observe any advantage
\cite{katzgraber:14,katzgraber:15}. Efforts have recently been focused on synthetic
benchmark problems constructed using planting methods
\cite{hen:15a,king:17}, post-selection
\cite{katzgraber:15,zhu:16,marshall:16}, and gadgets
\cite{denchev:16,mandra:18}. Despite all these efforts, only
Ref.~\cite{mandra:18} has demonstrated a {\em constant} speedup of
quantum annealing over the best available heuristics for a synthetic benchmark. However, a
constant speedup can easily be overcome by better implementations of
the classical heuristics on better classical hardware
\cite{matsubara:18,aramon:18}.

Reference \cite{mandra:16b} performed a comprehensive analysis of
quantum annealing compared to multiple classical heuristics using
synthetic benchmark problems. Quantum annealing was
outperformed by parallel tempering Monte Carlo
\cite{geyer:91,hukushima:96} with isoenergetic cluster updates
\cite{houdayer:01,zhu:15b}, as well as the Hamze--de Freitas--Selby
heuristic \cite{hamze:04,selby:14}, and the hybrid cluster method
\cite{venturelli:15a}. Nevertheless, within the class of {\em sequential algorithms},
quantum annealing on the D-Wave device as well as simulated
quantum annealing outperformed simulated annealing \cite{kirkpatrick:83}
and population annealing Monte Carlo
\cite{hukushima:03,machta:10,wang:15,wang:15e}, therefore being the most
efficient sequential optimization method to date. However, if the
best-known quantum inspired optimization methods
\cite{hartmann:01,juenger:01,hartmann:04,mandra:16b,mandra:17a,mandra:18}
are included, a different picture emerges. We emphasize that this statement concerns standard quantum annealing with a simple transverse field driver and a simple annealing schedule. 

Thus, a definite quantum speedup remains to be found, and it is an important goal of this article to present possible pathways for quantum annealing using advanced control parameters and more complex drivers that may point to a possible future direction toward unlocking the long sought-after quantum speedup over classical hardware.

\subsection{Role of entanglement in quantum annealing}

For various quantum technologies, especially quantum cryptography \cite{gisin2002quantum} and quantum-enhanced metrology \cite{pezze2018quantum}, entanglement has often been regarded as a resource that enables tasks beyond the capabilities of classical devices. 

In the case of quantum annealers, the presence of entanglement has been witnessed \cite{Lanting2017,albash2015reexamining}, but these experiments did not show any direct connection to the success probability or any other figure of merit of the quantum annealer. 
As several theoretical investigations studying different measures for entanglement indicate, the maximal entanglement achieved during the sweep has no implication on the success probability or a quantum speedup~\cite{batle2016multipartite}. 
However, Ref.~\cite{Hauke2015} found that the entanglement that survives at the end of the sweep does give an upper bound on the success probability. That is, if the system does not manage to get rid of the entanglement, the final state cannot assume a unique classical configuration, and the probability to reach the Ising ground state is diminished. 
A second indication for the role of entanglement in quantum annealing comes from Ref.~\cite{Bauer2015}, where it was shown that matrix product states with a larger cutoff for the allowed entanglement (as measured by the bond dimension) perform better in a simulation of the annealing procedure. 
Nevertheless, much further work is needed in order to understand the role of entanglement and quantumness towards any quantum speedup in a quantum annealing device.

\section{Perspectives on Methods\label{sec:perspectives_methods}}

Due to the difficulty of benchmarking quantum annealers for spin-glass problems numerically and experimentally, the ultimate gain possible by these quantum devices over classical machines is an unresolved question of current research \cite{ronnow2014defining,Katzgraber2018}. There are, however, various pathways towards boosting their performance, on the conceptual as well as on the experimental side. These represent the bulk of the rest of this article. This section treats perspectives on methods while those on physical implementations are discussed in Sec.~\ref{sec:implementations}. 

\subsection{Improvements by non-traditional terms and parameter control}

This section explains recent approaches to the improvement of the performance of quantum annealing, which use methods that lie out of the traditional formulation of quantum annealing based on the control of a simple transverse field applied uniformly over all qubits.

\subsubsection{Non-stoquastic Hamiltonians. }
\label{sec:Non-stoquastic}

Most theoretical and experimental studies of quantum annealing refer to the transverse-field Ising model with the Hamiltonian
\begin{equation}
H=A(t)H_0-B(t) \sum_i \sigma_i^x,
\label{stoqH}
\end{equation}
where $H_0$ is the classical Ising model of Eq.~(\ref{eq:hgk_sg}). 
The Hamiltonian in Eq.~(\ref{stoqH}) with non-negative $B(t)$ has all its off-diagonal elements real and non-positive on the standard basis to diagonalize $\sigma_i^z~(\forall i)$ (the computational basis). A Hamiltonian satisfying this condition is called `stoquastic', a compound of `stochastic' and `quantum' \cite{Bravyi2004}.\setcounter{footnote}{0}
\footnote{
More precisely, a Hamiltonian is stoquastic if the off-diagonal elements can be chosen real and non-positive in a basis which is a product of local bases \cite{Klassen2018}. The existence/absence of such a basis is not necessarily trivial \cite{Klassen2018,Marvian2018}.}

A stoquastic Hamiltonian can generally be simulated efficiently on a classical computer by the standard method of Trotter decomposition (see \cite{Hastings2013,Jarret2016} for exceptions), which leads to non-negative effective local Boltzmann factors \cite{LandauBinder2000}. This means that there is no sign problem, which plagues classical simulations of many non-trivial quantum-mechanical systems \cite{Loh1990,Troyer2005,LandauBinder2000}.  Another interesting property of a stoquastic Hamiltonian is that all coefficients of the ground-state wave function can be chosen to be non-negative in the standard basis, due to the Perron--Frobenius theorem, which excludes quantum interference effects in the wave function. These facts suggest the possibility that a stoquastic Hamiltonian may be devoid of essential quantum effects necessary for quantum advantage over classical algorithms. Related is the absence of a formal proof that the computational power of quantum annealing based on stoquastic Hamiltonians exceeds that of classical algorithms for specific problems with a few exceptions including the glued-tree problem \cite{Somma2012}, in which one ingeniously takes advantage of intermediate diabatic transitions.  See also \cite{Fujii2018}, where the non-standard adaptive measurements in the final state are shown to be useful to achieve quantum enhancement in stoquastic Hamiltonians. This latter development is worth further scrutiny to clarify the power of stoquastic Hamiltonians.

A non-stoquastic Hamiltonian has arbitrary signs (and even complex values) in off-diagonal elements in the standard basis.  A typical example is the Hamiltonian
\begin{equation}
H=-\sum_{i,j}J_{ij}\sigma_i^z \sigma_j^z -\sum_i h_i\sigma_i^z -\sum_i \Gamma_i\sigma_i^x +\sum_{i,j}\gamma_{ij}\sigma_i^x\sigma_j^x,
\label{nonstoqH}
\end{equation}
where $\Gamma_i\ge 0$ and some of the coefficients $\gamma_{ij}$ are positive
\footnote{
The local rotation $\sigma_i^x\to -\sigma_i^x, \sigma_i^y\to -\sigma_i^y, \sigma_i^z\to\sigma_i^z$ at some of the qubits $i$ can change the sign of $\Gamma_i$ and $\gamma_{ij}$. The Hamiltonian is non-stoquastic if some of $-\Gamma_i$ and $\gamma_{ij}$ remain positive under any such transformations.}.  Also, quantum annealing based on a non-stoquastic Hamiltonian of the form of Eq.~(\ref{nonstoqH}) is known to be equivalent to  gate-based quantum computation \cite{Biamonte2008}. See also Ref.~\cite{Aharonov2008}.

We are thus motivated to study both theoretically and experimentally the properties of non-stoquastic Hamiltonians. There exist, however, many obstacles because, for example, it is difficult to study numerically large-size spin-glass-like systems by classical simulations due to a large relaxation time even in the absence of the sign problem, see Sec.~\ref{sec:optimization problems}. Experimentally, 
realizing strong, tunable transverse interactions is also challenging, particularly for artificial atoms that are engineered to emulate spin-1/2 systems. For example, in the case of superconducting qubits (see Sec.~\ref{sec:Superconducting qubits}) using inductive couplings, the engineered magnetic moments $I_{x,i}$ used to realize the $\gamma_{ij}\sigma_i^x\sigma_j^x$ term are themselves dependent on the external control fields applied during the annealing protocol. This can be contrasted with an ideal spin-1/2 system, for which the magnetic moment $\mu_{x,z}$ is independent of the externally-applied magnetic fields $B_{x,y,z}$. Even when the $I_{x,i}$ are appropriately compensated, the magnitude of the resulting engineered coupling may not be as large as desired without increased engineering complexity, for example, through additional coupling elements to enhance the maximal value of $\gamma_{ij}$.

Nevertheless, several theoretical results have been reported that reveal enhanced computational capabilities of non-stoquastic Hamiltonians. In Refs.~\cite{Seki2012,Seoane2012,Seki2015,Nishimori2017}, it was shown that, in simple mean-field-type models, an additional $XX$ term with positive coefficient as in Eq.~(\ref{nonstoqH}) reduces first-order phase transitions present in the conventional transverse-field Ising model of Eq.~(\ref{stoqH}) to a second order transition.  As described in Sec.~\ref{subsection:energy_gap}, this change of the order of the phase transition means an exponential speedup relative to the stoquastic case.

\begin{figure} %
\centering
\includegraphics[width=0.9\textwidth]{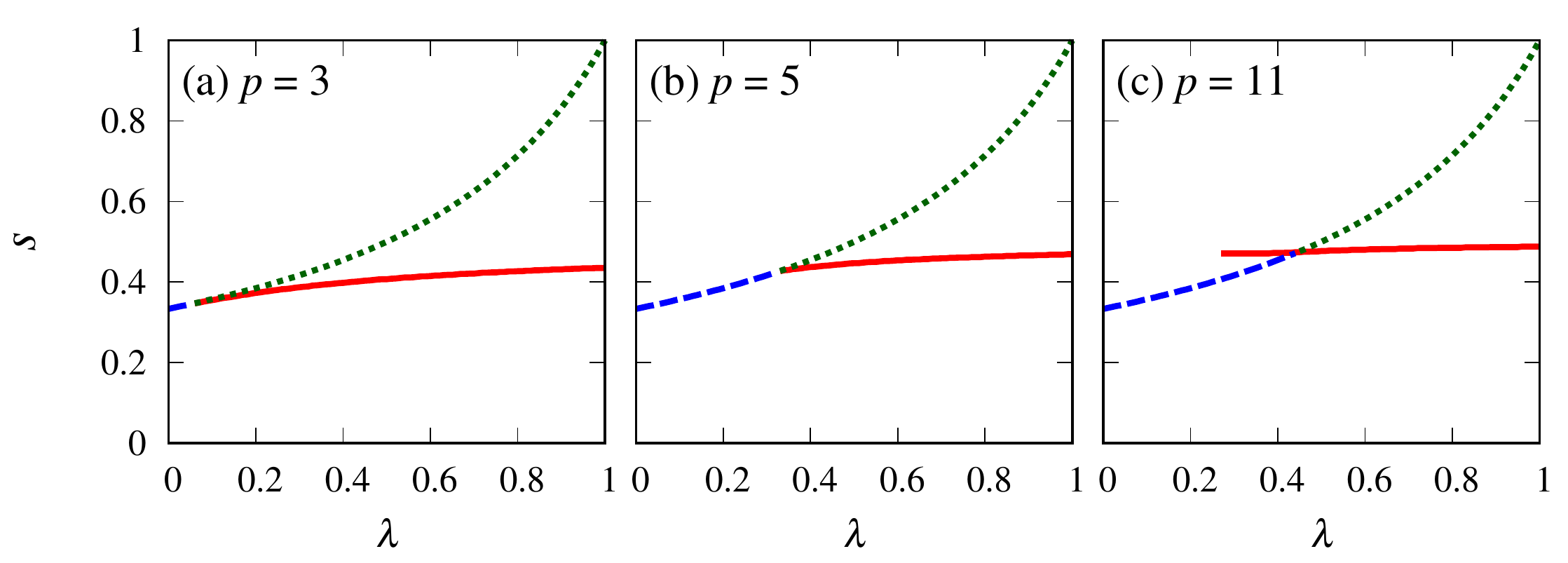}
\caption{Phase diagram of the non-stoquastic $p$-spin model with (a) $p=3$, (b) $p=5$, and (c) $p=11$. The red lines represent first-order transitions and the blue dashed lines are for second-order transitions.  The green dotted lines above the red lines are extensions of the blue dashed lines and do not directly represent phase transitions. The small portion of second-order transitions for $p=3$ near $\lambda=0$ comes from finite-size effects \cite{Durkin2018}. Based on \cite{Seki2012}.} 
\label{fig:p-spin_phase_diagram}
\end{figure}

More specifically, let us adopt the $p$-spin model as the cost function,
\begin{equation}
    H_0=-N\left(\frac{1}{N}\sum_{i=1}^N \sigma_i^z\right)^p,
    \label{eq:e-spin}
\end{equation}
where $p \ge 3$ is an integer. Although the ground state is trivial for this model---all spin up $\sigma_i^z=1$ (and all spin down $\sigma_i^z=-1$ for $p$ even)---the system undergoes a first-order phase transition if we use the conventional transverse field as the quantum driving Hamiltonian $H_1$.  This means that quantum annealing cannot solve this simple problem \cite{Jorg2010}.
It is nevertheless possible to avoid this difficulty by the introduction of an $XX$ term with a positive coefficient \cite{Seki2012},
\begin{equation}
  \!\!\!\!\! \!\!\!\!\!\! H(s,\lambda)=-s\lambda N\left(\frac{1}{N}\sum_{i=1}^N \sigma_i^z\right)^p-(1-s)\sum_{i=1}^N \sigma_i^x
    +s(1-\lambda)N\left(\frac{1}{N}\sum_{i=1}^N \sigma_i^x \right)^2,
    \label{eq:nonstoq-p}
\end{equation}
where $s$ and $\lambda$ are time-dependent parameters that change from $s=0$, $\lambda$ arbitrary at $t=0$ to $s=\lambda=1$ at $t=\tau$. The last term in Eq.~(\ref{eq:nonstoq-p}) makes the Hamiltonian non-stoquastic in the standard computational basis.  Conventional quantum annealing is recovered for $\lambda=1$, and $\lambda<1$ represents the introduction of non-stoquastic effects.  The phase diagram derived from statistical-mechanical calculations is displayed in Fig.~\ref{fig:p-spin_phase_diagram}.
This figure shows that the line of first-order transitions (red solid line) terminates in the middle of the phase diagram and is replaced by a line of second-order transitions (blue dashed). It is thus possible to choose a path connecting the starting point $s=0$, $\lambda$ arbitrary and the final point $s=\lambda=1$ without crossing a first-order transition. This means an exponential speedup relative to the conventional method that runs along the line $\lambda=1$. Note that this speedup is not against best classical algorithms and also that an ingenious trick enables one to simulate classically this particular non-stoquastic Hamiltonian \cite{Ohzeki2017}. These notes notwithstanding, it is encouraging that an example exists that explicitly shows exponential performance enhancement by non-stoquastic Hamiltonians relative to the stoquastic counterpart. 
Also, a small-size spin glass problem was shown to have enhanced performance by a non-stoquastic Hamiltonian \cite{Hormozi2017}.

Though some efforts have begun to understand how non-stoquastic effects lead to enhanced performance \cite{Albash2018}, our knowledge is still primitive and more extensive studies should follow to reveal the computational power of non-stoquastic Hamiltonians, which may lead to essential quantum-mechanical enhancement even against classical algorithms.

\subsubsection{Inhomogeneous driving of the transverse field. }
\label{sec:Inhomogeneous driving}

In conventional quantum annealing, the amplitude of a transverse field is controlled uniformly (homogeneously) over all qubits as in Eq.~(\ref{stoqH}).  As discussed in Sec.~\ref{subsection:energy_gap}, an adiabatic control of the homogeneous field from the initial value $B(0)=1$ to $B(\tau)=0$ causes a phase transition, which is one of the sources of difficulties.  A control of the field that is inhomogeneous over space, 
\begin{equation}
H=A(t)H_0-B(t)\sum_i \Gamma_i \sigma_i^x,
\label{eq:H_inhomogenoues1}
\end{equation}
where the amplitude $\Gamma_i$ depends on $i$,
may weaken the effects of a phase transition, because the number of qubits with the right critical values of coefficients for a phase transition to happen, $A(t)/(B(t)\Gamma_i)={\rm critical}$, is reduced from macroscopic $N$ for the uniform case ($\Gamma_i=1~\forall i$) to a much smaller number. In other words, only a small portion of the system is at the phase transition point, and the effects of phase transition may be weakened. This is the basic idea of inhomogeneous driving of the transverse field.
 
References \cite{Zurek2008,Dziarmaga2010,rams2016inhomogeneous,Mohseni2018} studied the one-dimensional Ising model with an inhomogeneous field analytically based on the Kibble--Zurek mechanism \cite{Kibble1976,Zurek1985} and found evidence for better performance as measured by the number of defects (misaligned spins) in the final state.  Numerical \cite{farhi2011} and analytical \cite{Dickson2011,Dickson2012} work on typical combinatorial optimization problems of 3-SAT and maximum independent set problems also indicated improvements by inhomogeneity of the transverse field.  Equilibrium statistical mechanical analysis of the $p$-spin model showed that the first-order phase transition is removed by the inhomogeneity of the transverse field \cite{Susa2018a,Susa2018b}, which means an exponential speedup over the homogeneous protocol.

\begin{figure} %
\centering
\includegraphics[width=0.38\textwidth]{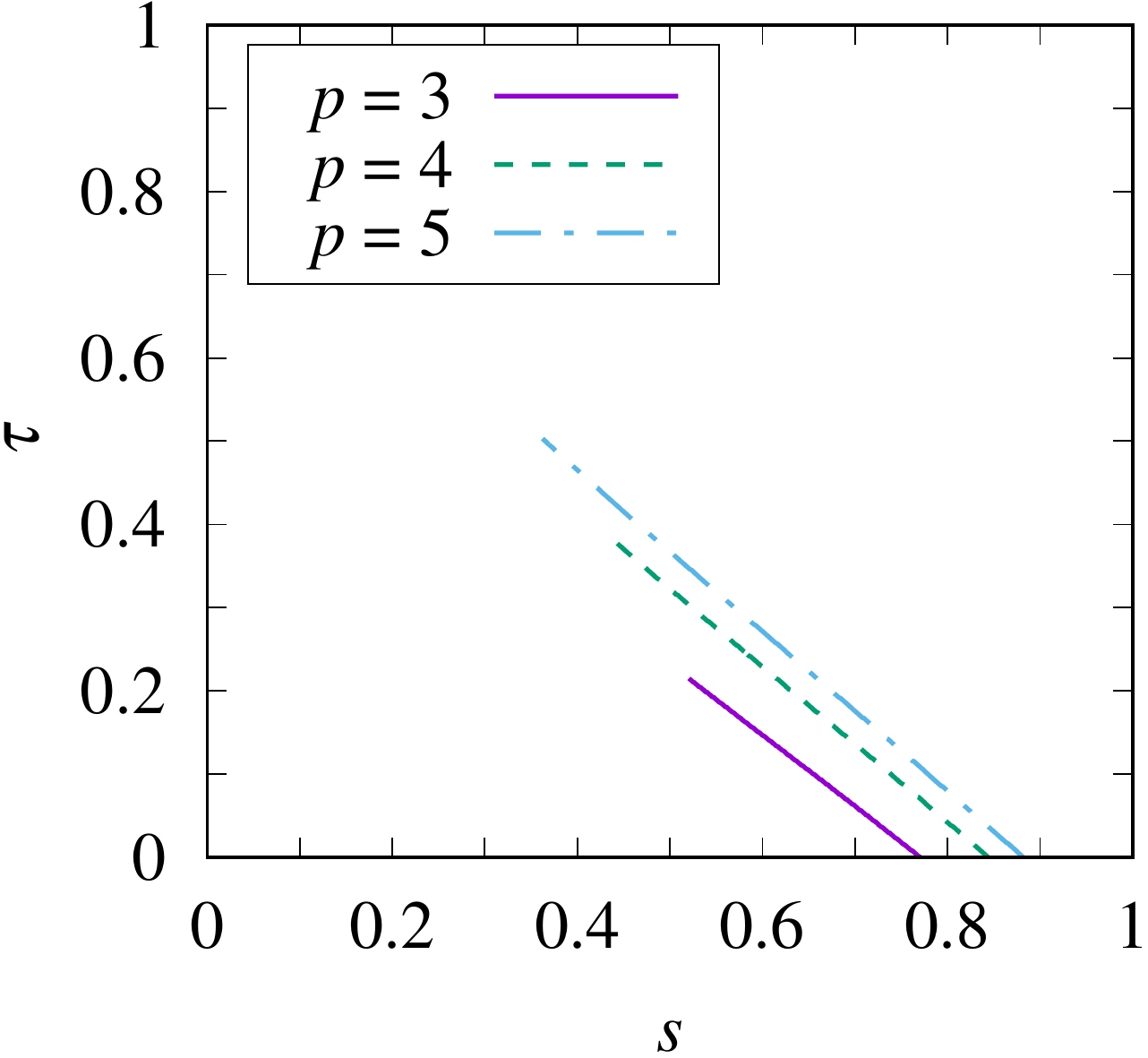}
\caption{Phase diagram for the $p$-spin model under inhomogeneous field. Each line represents a series of first-order transitions for a given $p$, which extends to the middle of the phase diagram from the axis $\tau=0$ corresponding to the conventional homogeneous field. Taken from \cite{Susa2018b}.} 
\label{fig:inhomogeneous}
\end{figure}

Following Refs. \cite{Susa2018a,Susa2018b}, let us choose in Eq.~(\ref{eq:H_inhomogenoues1}) the coefficients as $A(t)=t/\tau (\equiv s), B(t)=1$, and the amplitude $\Gamma_i=1$ for $i=1, 2,\cdots, N(\tau-1)~(0\le \tau\le 1)$ for qubits with field and $\Gamma_i=0$ for $i=N(\tau-1)+1,\cdots, N$ for qubits without field, i.e.,
\begin{equation}
    H=sH_0-\sum_{i=1}^{N(1-\tau)}\sigma_i^x ,
    \label{eq:H_inhomogenoues2}
\end{equation}
where $H_0$ is the cost function of the $p$-spin model Eq.~(\ref{eq:e-spin}).  Here, the parameter $\tau$ is for the ratio of qubits without transverse field (not to be confused with the computation time), and we start the annealing process with $s=\tau=0~(H=-\sum_{i=1}^N\sigma_i^x)$ and end with $s(>0)$ arbitrary and $\tau=1$. The phase diagram is drawn in terms of these two parameters as in Fig.~\ref{fig:inhomogeneous}, where $\tau=0$ represents the conventional protocol with transverse field homogeneously applied to all qubits.
Each line represents a series of first-order transitions for a given $p$, starting from the $\tau=0$ line and ending in the middle of the phase diagram. It is clearly possible to choose a path connecting the starting point $s=\tau=0$ and ending at $s$ arbitrary and $\tau=1$, implying the existence of a way to switch off the field qubit by qubit as time $s$ proceeds without hitting a phase transition. This means an exponential speedup as compared to the conventional method ($\tau=0$) in the same sense as in the previous section for non-stoquastic Hamiltonians.

Experimentally, the current version of the D-Wave device features the option of an `anneal offset,' which is to change the amplitude of the transverse field individually to some extent.  Although it is not a complete realization of inhomogeneous field driving as described above, it nevertheless shows improved performance for some hard optimization problems \cite{Adame2018}.
In addition, it improves the biased sampling \cite{mandra:17} that has been observed with standard transverse-field annealing protocols \cite{Lanting2017}.

\subsubsection{Reverse annealing. }

Reverse annealing is another protocol to control the amplitude of the transverse field in a non-traditional way.  As proposed in Ref.~\cite{Perdomo-Ortiz2010} (see also \cite{Chancellor2017}), we may be able to take advantage of  partial knowledge on the correct solution, and use it as a candidate state expected to be close to the solution. Such a candidate may have been obtained by another method like classical simulations. We then start from this specific classical state under zero transverse field and increase the amplitude of the transverse field to a pre-assigned finite value and then decrease it again to zero. The system explores the space of states near the candidate state, which may be more efficient than the whole-space search from the {\em tabula rasa} initial condition in the traditional method.

\begin{figure*}[thb]
\centering
    \begin{tabular}{c}
      \begin{minipage}{0.32\linewidth}
			\centering
          \includegraphics[width=5cm,clip]{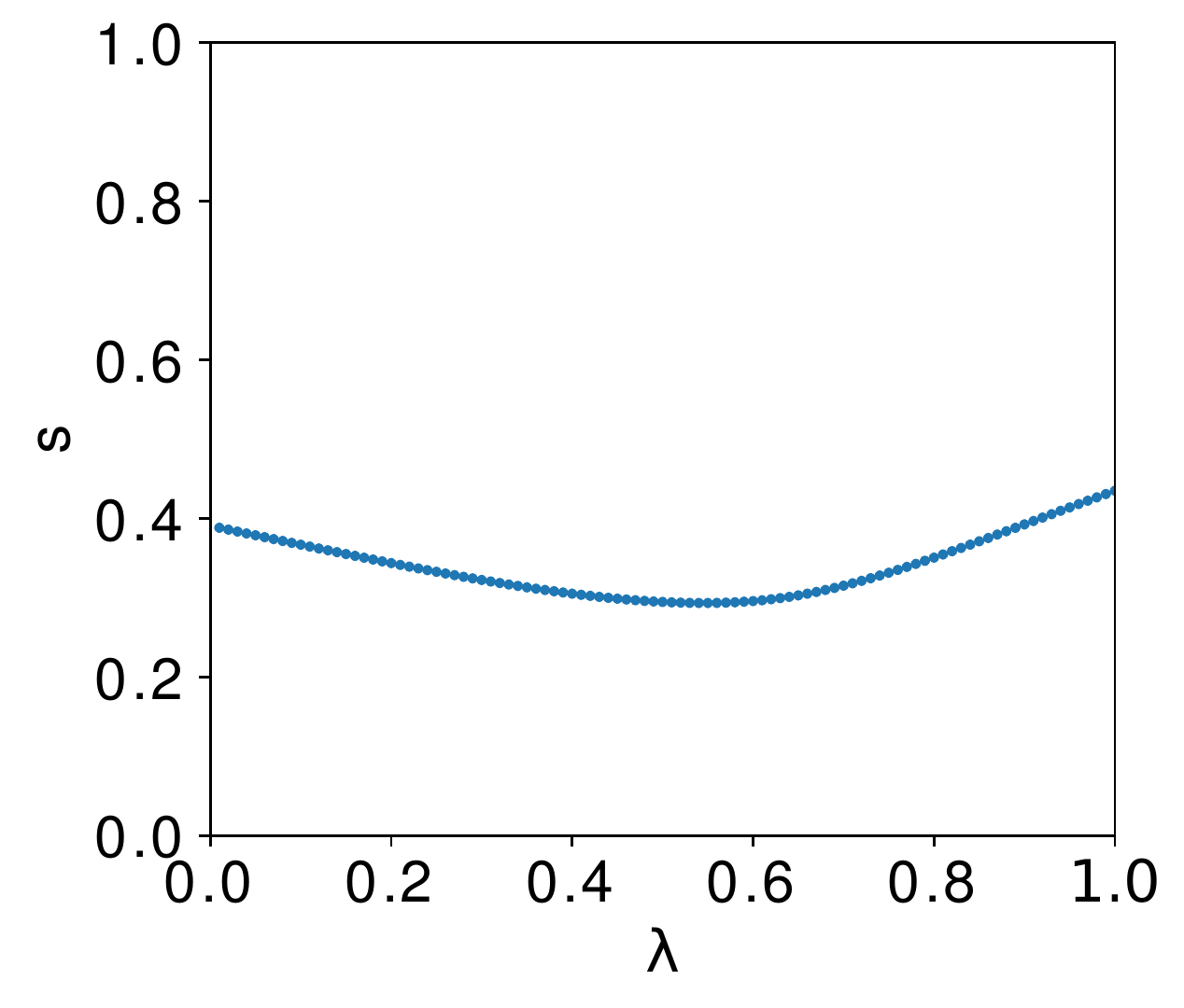}
          \hspace{12mm} (a)$\;c=0.7$
      \end{minipage}
      \begin{minipage}{0.32\linewidth}
        \begin{center}
          \includegraphics[width=5cm,clip]{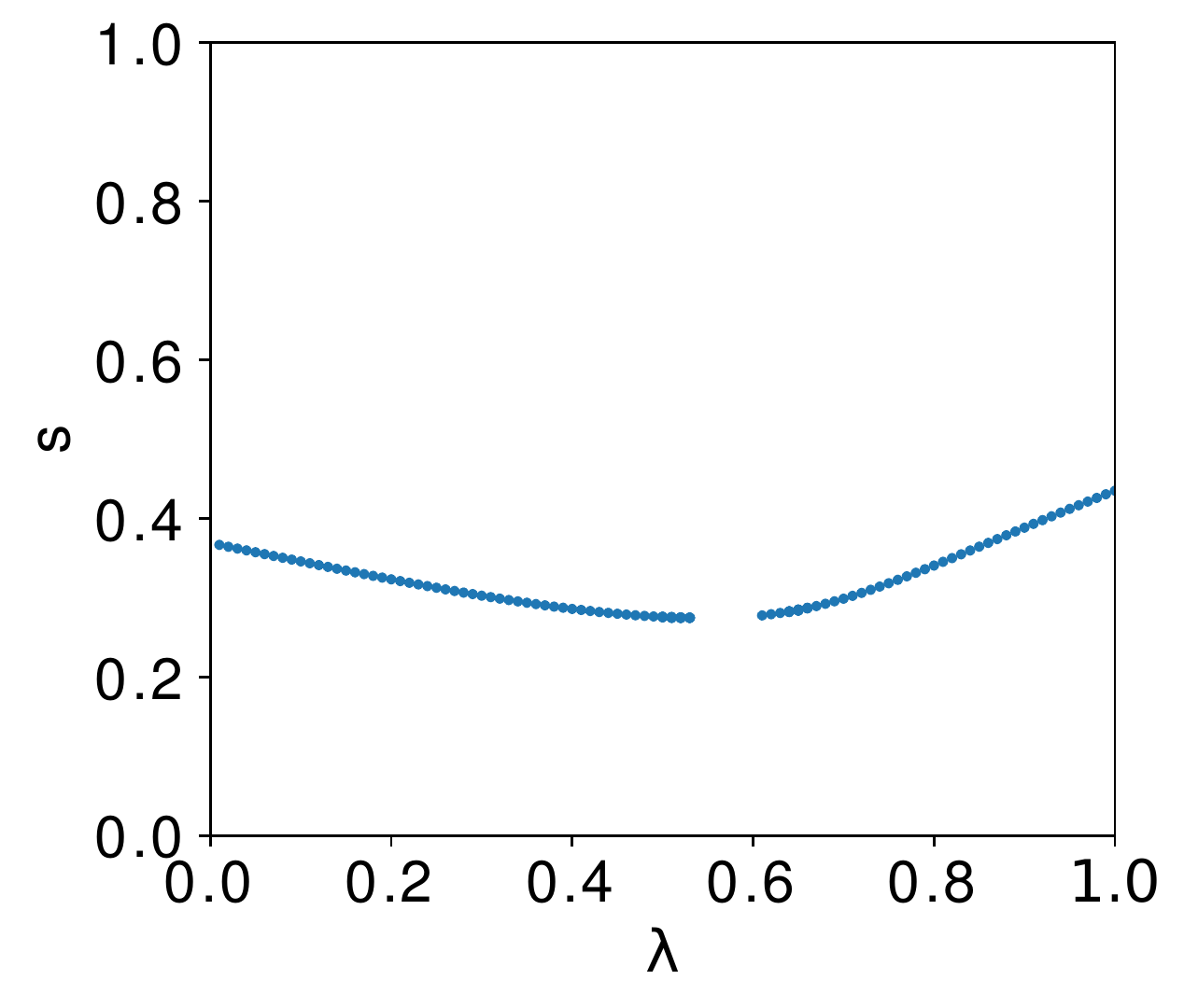}
          \hspace{10mm} (b)$\;c=0.74$
        \end{center}
      \end{minipage}
	\begin{minipage}{0.32\linewidth}
			\centering
          \includegraphics[width=5cm,clip]{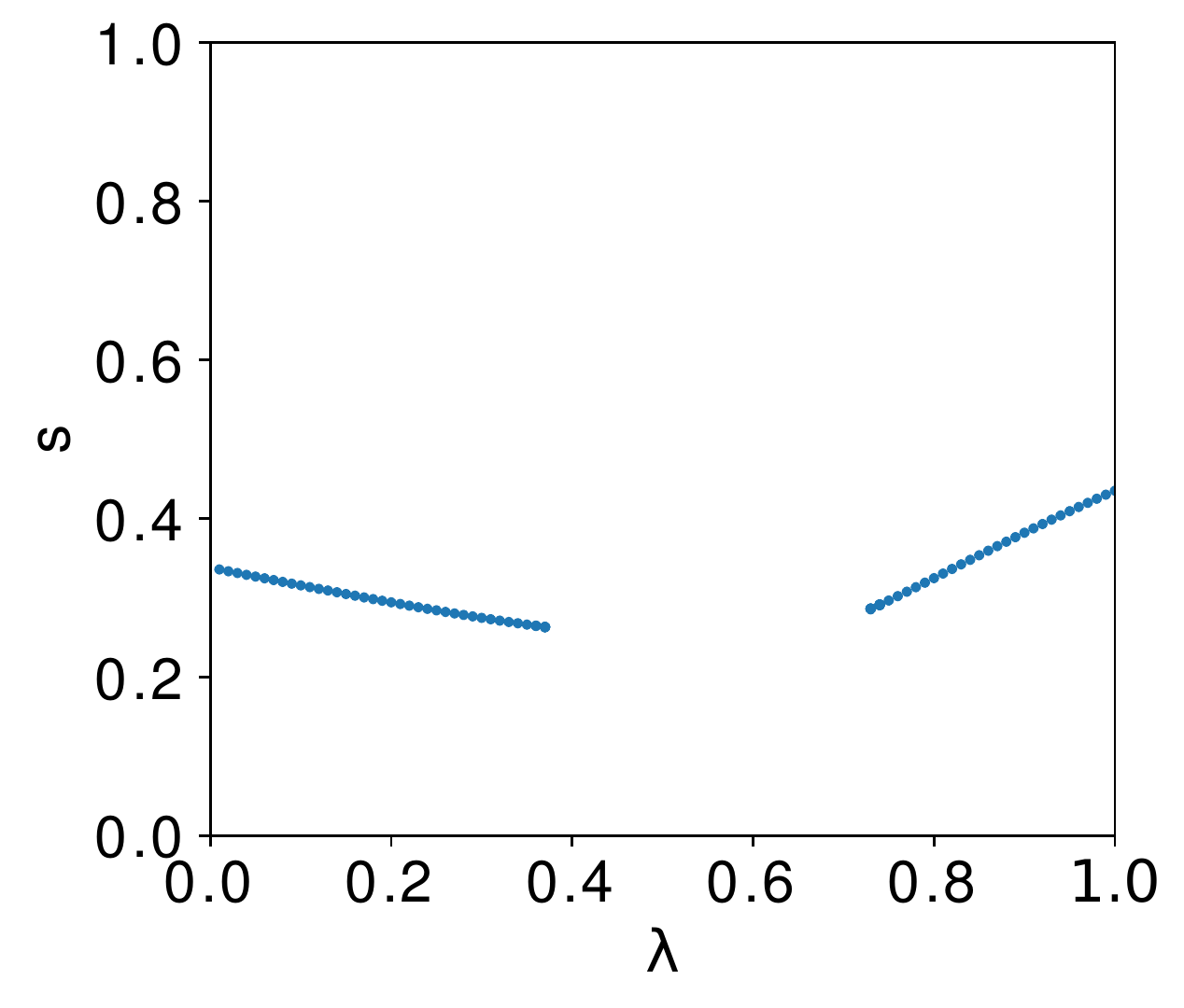}
          \hspace{10mm} (c)$\;c=0.8$
      \end{minipage}
    \end{tabular}
    \caption{Phase diagrams on the $\lambda$-$s$ plane for $p=3$ and three values of $c$.  The blue curves represent lines of first-order transitions. A path exists in (b) and (c) to connect the initial state at $s=\lambda=0$ and the final state at $s=\lambda=1$ without hitting a first order transition.  Taken from \cite{Ohkuwa2018}.}
\label{fig:phase_diagram_reverse_annealing}
\end{figure*}

Numerical results on a typical combinatorial optimization problem of 3-SAT indicate encouraging results \cite{Perdomo-Ortiz2010}. An analysis of the $p$-spin model revealed that a first-order phase transition in the traditional formulation disappears by reverse annealing if the initial condition is chosen close to the correct solution \cite{Ohkuwa2018}.
In order to enforce the candidate state as the initial state, we add a term to the Hamiltonian of the $p$-spin model
\begin{equation}
    H=sH_0-(1-s)\lambda \sum_{i=1}^N\sigma_i^x -(1-s)(1-\lambda)\sum_{i=1}^N \epsilon_i\sigma_i^z,
    \label{eq:reverse_annealing}
\end{equation}
where $s$ and $\lambda$ are parameters to control the time evolution, running from $s=\lambda=0$ initially to $s=\lambda=1$ finally. The parameter $\epsilon_i(=\pm 1)$ represents the candidate state, and the final term in the Hamiltonian Eq.~(\ref{eq:reverse_annealing}) constrains the initial state to $\sigma_i^z=\epsilon_i~(\forall i)$ at $s=\lambda=0$.  The ground state is trivial  for the $p$-spin model ($\sigma_i^z=1~(\forall i)$ for $p$ odd and also $\sigma_i^z=-1~(\forall i)$ for $p$ even), and $\epsilon_i$ is expected to be $1$ with a probability $c$ reasonably close to 1. In the phase diagram depicted in Fig.~\ref{fig:phase_diagram_reverse_annealing}, the conventional quantum annealing (along the vertical line $\lambda=1$) is seen to encounter a first order transition (blue curve) during the time evolution along the $s$ axis.
The line of first order transitions drawn blue terminates in the middle of the phase diagram if the initial condition is fairly close to the final solution, i.e. $c$ close to 1 as in panels (b) and (c). This signifies an exponential speedup over the conventional approach along the line $\lambda=1$.

Further theoretical, numerical, and experimental studies would help us understand when and how reverse annealing is effective.

\subsection{Non-adiabatic schemes }
\label{subsec:non-adiabatic}

As discussed in Sec.~\ref{sec:adiabatic_approximation}, the adiabatic condition poses a fundamental speed limit to adiabatic quantum computing given by near-crossings of energy levels. A strategy to overcome this fundamental limitation is to detach from the adiabatic approximation and introduce non-adiabatic methods, so-called  counter-diabatic driving \cite{demirplak2003adiabatic,demirplak2005assisted,berry2009transitionless,Polkovnikov,Takahashi2013} and/or short-cut to adiabaticity \cite{del2013shortcuts,deffner2014classical,torrontegui2013shortcuts}. Let us summarize the main idea of counter-diabatic driving. 

We consider the frame rotating with respect to a unitary operator $U$ and define the Hamiltonian in the rotating frame $\hat{\mathcal{H}}=U^{\dagger} \mathcal{H} U$ and the corresponding wave function $\ket{\hat{\psi}}=U^{\dagger} \ket{\psi}$. The Schr\"odinger equation is then 
\begin{eqnarray}
\hat{\mathcal{H}} \ket{\hat{\psi}}&=& i \hbar \partial_t \ket{\hat{\psi}} =i \hbar \partial_t(U^{\dagger} \ket{\psi}) \nonumber \\
&=&i \hbar \partial_t \lambda \, \partial_{\lambda} U^{\dagger} \ket{\psi}  + U^{\dagger} \mathcal{H} \ket{\psi}  \nonumber \\
&=&\partial_t \lambda \, (i \hbar \partial_{\lambda} U^{\dagger} U) \ket{\hat{\psi}} + U^{\dagger} \mathcal{H} U \ket{\hat{\psi}} = (\hat{\mathcal{H}} - \dot{\lambda} \mathcal{A}_{\lambda}) \ket{\hat{\psi}} \label{eq:schrodingerrot}.
\end{eqnarray}
where have defined the adiabatic gauge potential as
\begin{equation}
\mathcal{A}_{\lambda}= i \hbar U^{\dagger} \partial_{\lambda}U. \label{eq:gauge}
\end{equation}
In the laboratory frame the gauge potential eliminates the off-diagonal terms of the moving Hamiltonian and thus any transition. This can be seen from differentiating $\hat{\mathcal{H}}(\lambda)$ with respect to $\lambda$. We obtain
\begin{equation}
\partial_{\lambda}\hat{\mathcal{H}}=U^{\dagger} \partial_{\lambda} \mathcal{H} U+\frac{i}{\hbar}[\mathcal{A}_{\lambda},\hat{\mathcal{H}}]. \label{eq:eqA3}
\end{equation} as $[\partial_{\lambda} \hat{\mathcal{H}}, \hat{\mathcal{H}}]=0$.  
The adiabatic gauge potential $\mathcal{A}_{\lambda}$ from this equation is an exact formulation of the diabatic transition due to a general unitary transformation. The idea of counter diabatic driving is to cancel this term by introducing it with a negative sign in the Hamiltonian 
\begin{equation}
H_{\textrm{CD}}(t)=H(t)+\dot{\lambda} \mathcal{A}_{\lambda}, \label{eq:hcd}
\end{equation}
Here, $H(t)$ is the original time-dependent Hamiltonian from the adiabatic quantum computing protocol. In adding the exact counter-diabatic gauge potential, transitions get suppressed and the system remains in its instantaneous ground state for all times and sweeps \cite{demirplak2005assisted}.

However, from an experimental point of view, the exact gauge potential $\mathcal{A}_{\lambda}$ is of limited use, because the terms that appear in the Hamiltonian are highly non-local and include all orders of $k$-body interactions \cite{demirplak2005assisted,Polkovnikov}. It is thus a natural question whether one can find approximate counter-diabatic Hamiltonians that can be built with the available resources and still sufficiently improve the ground-state fidelity. 

Recently, a variational method to find approximate counter-diabatic Hamiltonians has been introduced by Sels and Polknovnikov \cite{Polkovnikov}. In this scheme, the recipe to construct approximate counter-diabatic terms is to use an ansatz that includes the experimental available interactions, which is then variationally optimized with respect to the adiabatic gauge potential. Here, we introduce $\mathcal{A}^*_{\lambda}$ as the approximate Gauge potential which we variationally optimize. The idea is to minimize the operator distance between the exact Gauge potential $\mathcal{A}_{\lambda}$ and the approximate ansatz Gauge potential $\mathcal{A}^*_{\lambda}$ defined as
\begin{equation}
D^2(\mathcal{A}^*_{\lambda})=Tr[(G_{\lambda}(\mathcal{A}_{\lambda})-G_{\lambda}(\mathcal{A}^*_{\lambda}))^2]
\end{equation}
where $G$ is the Hilbert-Schmidt norm 
\begin{equation}
G_{\lambda}(\mathcal{A}^*_{\lambda})=\partial_{\lambda} H(t)+i[\mathcal{A}^*_{\lambda},H(t)].\label{eq:hsnorm}
\end{equation}
The aim is to minimize the operator distance with respect to the variational parameters of $\mathcal{A}^*_{\lambda}$. In Ref. \cite{Polkovnikov} it was shown that operator distance is equivalent to the minimum of the action 
\begin{equation}
\mathcal{S}(\mathcal{A}^*_{\lambda})=Tr[G^2_{\lambda}(\mathcal{A}^*_{\lambda})] \label{eq:eq6}
\end{equation}
associated with the approximate adiabatic gauge potential $\mathcal{A}^*_{\lambda}$, i.e.,
\begin{equation}
\frac{\delta \mathcal{S}(\mathcal{A}^*_{\lambda})}{\delta \mathcal{A}^*_{\lambda}}=0, \label{eq:eq7}
\end{equation}
where $\delta$ denotes the partial derivative.

In Ref.~\cite{Polkovnikov}, this variational approach was demonstrated for a range of model systems, including an Ising spin chain. In this model, the counter-diabatic terms are derived from the ansatz
\begin{eqnarray}
\mathcal{A}^*_{\lambda} = \sum_j \left[ \alpha_j \sigma_j^y  + \beta_j \left(\sigma_j^y \sigma_{j+1}^z + \sigma_j^z \sigma_{j+1}^y \right) \right. \\
   \left. + \gamma_j \left( \sigma_j^y \sigma_{j+1}^x + \sigma_j^z \sigma_{j+1}^x\right) \right].
 \label{eq:cdspinchain}
\end{eqnarray}
Here, the switching functions $\alpha$, $\beta$ and $\gamma$ describes a smooth function of the form
\begin{equation}
\lambda(t)=\lambda_0+(\lambda_f-\lambda_0)\sin^2\left(\frac{\pi}{2}\sin^2\left(\frac{\pi t}{2 \tau}\right)\right), 
\label{eq:cdlambda}
\end{equation}
where $\tau$ is the sweep time and $\lambda_0$ and $\lambda_f$ the initial and finals values, respectively. This particular form of the switching function is chosen to satisfy the condition that first and second derivatives at $t=0$ and $t=\tau$ vanish. Adding a Hamiltonian term of the form of Eq.~(\ref{eq:cdspinchain}) with a switching function as in Eq.~(\ref{eq:cdlambda}) to the standard annealing Hamiltonian improves the ground-state fidelity considerably. However, Eq.~(\ref{eq:cdspinchain}) consists of all-to-all $\sigma^y \sigma^x$ couplings, which is experimentally challenging.

In Ref.~\cite{hartmann2018rapid}, the variational counter diabatic terms have been combined with the LHZ implementation of all-to-all spin models \cite{lechner2015quantum}. The LHZ architecture encodes the optimization problem in local fields, as compared to the conventional Ising spin representation where the problem is encoded in pair interactions. The LHZ Hamiltonian reads as 
\begin{equation}
\label{eq:LHZ_Hamiltonian}
H = A(t) \sum_i^K J_i \sigma_i^z + B(t) \sum_i^K \sigma_i^x + C(t)\sum_n^{K-N+1} C_n \sigma_i^z\sigma_j^z\sigma_k^z\sigma_l^z, \label{eq:lhz}
\end{equation}
where $A$, $B$, and $C$ are switching protocols, $K$ is the number of connections (e.g., $K=N(N-1)/2$ for all-to-all models) and the indices in the last term run over four qubits in each plaquette of a square lattice. 

The main result of Ref.~\cite{hartmann2018rapid} is a protocol to find optimal counter-diabatic terms with local qubit operations from a single-qubit ansatz for the counter-diabatic terms of the form  $\mathcal{A}_{\lambda}^*=\sum_i \alpha_i \sigma_i^y$. The complete protocol is a quantum-classical hybrid approach where a variational parameter is updated from the outcome of a measurement from a quantum protocol. This protocol consists of two distinct variational optimizations: First, the analytical result from the gauge potential and second, the iterative optimization of a single optimization parameter from repeated measurements. 

The possibility to loosen the strict adiabatic condition by using additional information about the system, be it variationally or analytically, may open the door to a wide range of theoretical and practical advances in adiabatic quantum computing. Approximate counter-diabatic Hamiltonians may allow one to trade sweep time for available information about the problem and experimentally available resources. 

\section{Perspectives on Implementations\label{sec:implementations}}

Further theoretical advances are just one missing piece in order to arrive at quantum annealers that solve problems of practical relevance. Just as important will be a strong continued progress in the design of quantum devices. In this section, we discuss the state of the art and future prospective of superconducting qubits, the paradigm platform for quantum annealing, as well as ideas for alternative platforms based on Rydberg atoms and trapped ions. 

\subsection{Superconducting qubits }
\label{sec:Superconducting qubits}

Superconducting qubits are artificial spin-$1/2$ systems comprising lithographically defined inductors, capacitors, wires, and Josephson tunnel junctions. Their lithographic scalability, compatibility with electrical control signals, potential for high coherence, and the existence of a superconducting classical digital logic all converge to make the superconducting qubit modality the most advanced quantum annealers available today. 

\subsubsection{Current state-of-the-art and limitations. }

The current state-of-the-art for quantum annealers is the \textit{D-Wave 2000Q}~\cite{Dwave2000q}, a quantum annealing system comprising in excess of 2000 superconducting qubits and controlled in part by superconducting digital logic, see Fig.~\ref{fig:Washington}. The D-Wave 2000Q uses a ``Chimera graph'' architecture \cite{bunyk:14}, built around an 8-qubit unit cell. Each qubit has an \textit{intra}-cell connectivity to four other qubits and \textit{inter}-cell connectivity to two other qubits in neighboring unit cells. Problems are embedded on the Chimera architecture using proprietary software, and as mentioned in the Introduction, numerous examples have been implemented on D-Wave quantum annealers to date. 

Despite D-Wave's truly remarkable engineering achievement, it remains unknown if the D-wave quantum annealer -- or quantum annealers in general -- will afford quantum advantage for a general class of problems. 
The limitations of contemporary superconducting quantum annealers include:
\begin{itemize}
\item Short coherence times (compared with gate-model quantum computers). 
\item Limited connectivity (thus limiting the optimization problems that can be solved, see Sec.~\ref{sec:optimization problems}, or requiring mappings such as the LHZ approach, Eq.~(\ref{eq:lhz})). 
\item Solely stoquastic (ZZ) coupling (thus excluding any potential advantage from non-stoquastic Hamiltonians, see Sec.~\ref{sec:Non-stoquastic}). 
\item Solely two-body interactions (hence precluding higher-order PUBOs, see Sec.~\ref{sec:optimization problems}).  
\item Limited ability to tailor the annealing schedule for different qubits (hence limiting inhomogeneous-driving schemes, see Sec.~\ref{sec:Inhomogeneous driving}).
\end{itemize}
At the time of this writing, it is not known if remedying these limitations will practically lead to quantum advantage, but addressing them is a good place to start. 

\subsubsection{Introduction to superconducting qubits. }

\begin{figure} %
\centering
\includegraphics[width=1.0\textwidth]{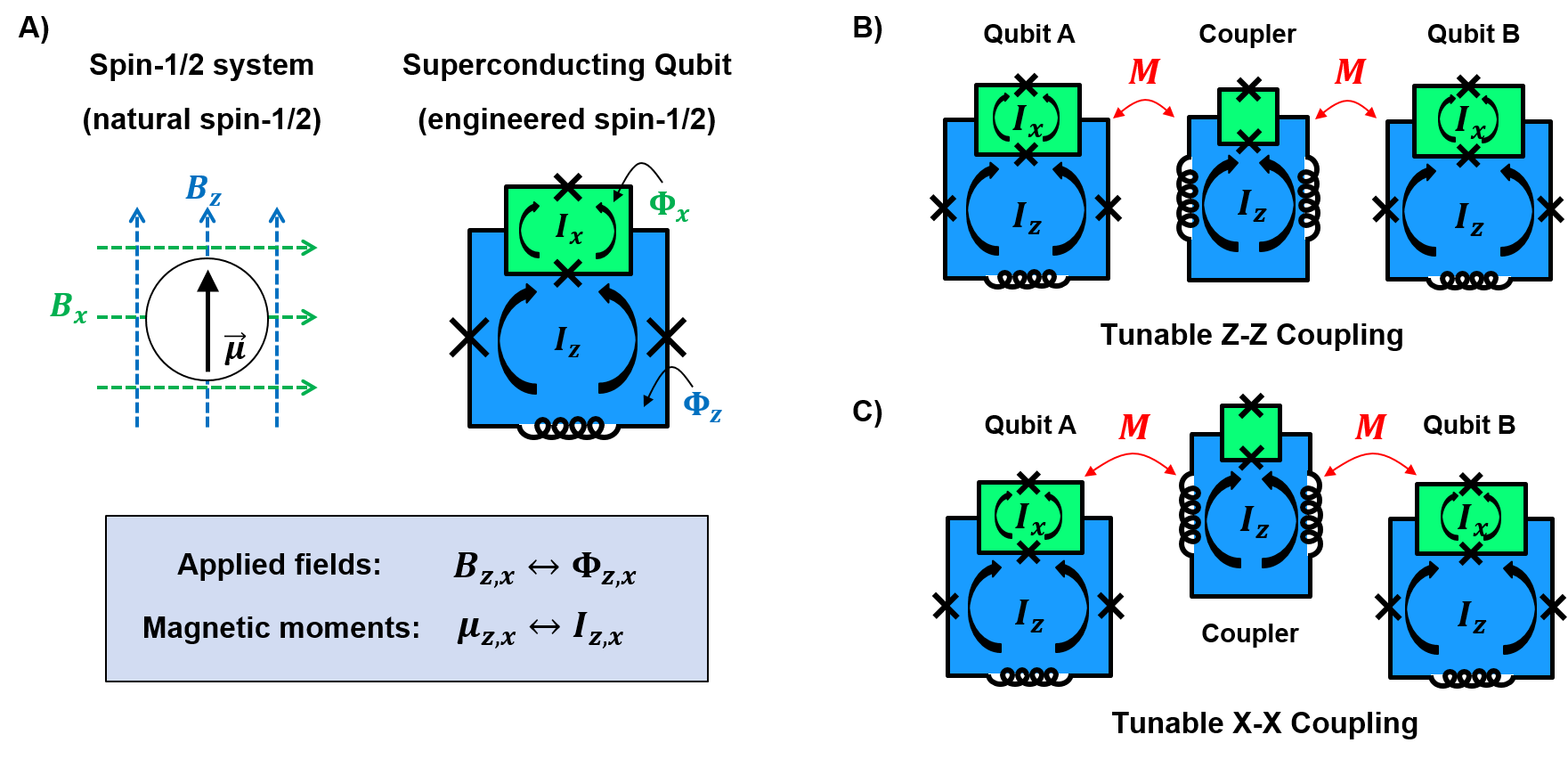}
\caption{Superconducting qubits. (A) Comparison of an ideal spin-1/2 system and an engineered emulation of a spin-1/2 using a superconducting qubit. (B) Tunable Z-Z coupling mediated by an RF SQUID coupler. (C) Tunable X-X coupling mediated by an RF SQUID. This type of X-X coupling has limitations due to a field-dependent magnetic moment of the qubit.} 
\label{fig:superconducting-qubits}
\end{figure}

As mentioned, superconducting qubits are lithographically defined circuits comprising inductors, capacitors, interconnects, and Josephson junctions~\cite{Oliver2013}. When cooled to milliKelvin temperatures, these circuits exhibit quantized energy levels corresponding to quantized states of electric charge or magnetic flux, depending on the parameter values of these various circuit elements. In the context of gate-model quantum computation, the lowest-two of these energy levels are generally used as a quantum bit to represent logical states $|0\rangle$ and $|1\rangle$. Here, coherence within this reduced manifold is a premium, and, over the past 20 years, research groups worldwide have devoted significant resources to improving qubit coherence through a combination of materials, fabrication, and design advances. State of the art for gate-model qubits stands at approximately 100 $\mu$s for both the energy decay time (longitudinal relaxation time $T_1$) and coherence time (transverse relaxation time $T_2$)~\cite{Paik2011,Rigetti2012,Jin2015,Yan2016,Yan2018}. 

The approaches taken to achieve this level of coherence are often incompatible with the needs for quantum annealing. For example, qubits with fixed frequencies (determined during fabrication) or qubits with only $\sigma^z$ tunability reduce the number of noise channels that can impact coherence, but do not allow for fully tunable transverse Ising Hamiltonians with $\sigma^z$ and $\sigma^x$ terms. Additionally, the materials (typically aluminum) and fabrication approaches (typically single-metal layers) are kept as simple and minimal as possible to avoid introducing unwanted defects that reduce $T_1$ and $T_2$, at the expense of circuit complexity. There are several types of gate-model qubits in common use today, including the transmon, capacitively shunted flux qubit, and fluxonium.

In contrast, the superconducting qubits used for quantum annealing are generally more complex by necessity. The most common type is called a ``X- and Z-tunable flux qubit,'' comprising superconducting loops interrupted by Josephson junctions (indicated by the symbol "X" in Fig.~\ref{fig:superconducting-qubits}A)~\cite{Orlando1999,Mooij1999}. In its conventional form, the $Z$-field is tunable by threading a magnetic field $\Phi_{\textrm{z}}$ through the $\sigma^z$-loop (blue). Clockwise and counterclockwise circulating currents are the two classical (diabatic) states of this circuit and serve to help screen $\Phi_{\textrm{z}}$. When $\Phi_{\textrm{z}}=\Phi_0/2$, where $\Phi_0=h/2e$ is the superconducting quantum unit of flux, both circulating-current states have the same energy and the system is frustrated. The green loop and its junctions serve to couple the diabatic currents, opening an avoided crossing at this degeneracy point. The size of this avoided crossing, the strength of the $X$-field, is tunable by a magnetic flux $\Phi_{\textrm{x}}$ applied to the $\sigma^x$ loop (green). 
In essence, the potential profile for this circuit is a two-dimensional double well. Each well corresponds to a circulating current, and the applied flux $\Phi_{\textrm{z}}$ tilts the wells. The wells are tunnel-coupled through a potential barrier, and the height of this barrier (and, thus, the tunneling rate) is tuned by the applied flux $\Phi_{\textrm{x}}$. 

Thus, a X- and Z-tunable flux qubit behaves as a spin-1/2 system in a magnetic field. The applied magnetic flux $\Phi_{\textrm{x,z}}$ and magnetic moment $I_{\textrm{x,z}}\times A_{\textrm{x,z}}$ of the flux qubit, where $A_{\textrm{x,z}}$ is the loop area, correspond to the applied field $B_{\textrm{x,z}}$ and magnetic moment $\mu_{\textrm{x,z}}$ of the natural spin-1/2, respectively. One important distinction, however, is that the magnetic moment $I_{\textrm{x}}$ of the flux qubit is generally \em{dependent }\em on the applied flux $\Phi_{\textrm{x}}$. This should be contrasted with an ideal spin-1/2, for which the magnetic moment is constant and independent of the applied field. This difference must be compensated to realize ideal spin-1/2 behavior by either a careful calibration of the $X$-field and, more likely, additional complexity to realize the full range and strength of desired $\sigma_i^x \sigma_j^x$ coupling. 

\subsubsection{Approaches to coupling. }

Qubit coupling is achieved using mutual inductance between qubits and the qubit circulating currents. Although a direct interaction between qubits is possible in principle, it is not useful in practice, as the coupling strength cannot be tuned independently of the single-qubit field strengths. Rather, achieving large and tunable $\sigma_i^z \sigma_j^z$ and $\sigma_i^x \sigma_j^x$ coupling is implemented using intermediate qubits called ``couplers'' (see Fig.~\ref{fig:superconducting-qubits}B, C). 

The couplers are qubits operated in a regime where they behave as tunable inductors with effective inductance $L_{\textrm{eff}}$. The X-field for the coupling is biased large enough that the coupler always remains in its ground state. Taking $\sigma_i^z \sigma_j^z$ coupling for example, the coupling energy $J$ between two qubits, each with circulating current $I_{\mathrm{z}}$ and mutual coupling $M$, is $J=(I_{\textrm{z}}M)^2/L_{\textrm{eff}}$. In turn, the inductance of the coupler is tunable using $\Phi_{\mathrm{z}}^{\mathrm{c}}$ applied to the coupler $Z$-loop, and the coupling can be tuned from ferromagnetic ($J<0$) to antiferromagnetic ($J>0$). For a detailed description of how this works, see Refs.~\cite{Harris2007,Weber2017}. A similar approach can be used to couple the $X$ loops and achieve $\sigma_i^x \sigma_j^x$. By tuning the sign of the $\sigma_i^x \sigma_j^x$ coupling, one can realize both stoquastic and non-stoquastic Hamiltonians. Furthermore, the couplers themselves can be cascaded to achieve fanout~\cite{Kerman2019}. We note that $\sigma_i^y \sigma_j^y$ coupling can also be realized using capacitive coupling. 

There is a fundamental trade-off between strong coupling and high coherence. Large circulating currents lead to strong coupling, since $J\propto I_{\textrm{x,z}}^2$. However, coherence is severely impacted by the size of the circulating current and its corresponding magnetic moment. For example, the energy relaxation time scales as $1/I_{\textrm{x,z}}^2$ and the dephasing time scales as $1/I_{\textrm{x,z}}$~\cite{Weber2017}. Thus, quantum annealers that use large circulating currents generally have low coherence times (around 10 ns), whereas gate-model qubits with high coherence (around 100 $\mu$s) all use small circulating currents~\cite{Yan2016}. The couplers can be used to compensate a weak coupling due to small circulating currents, and coupling strengths $J=1$ GHz have been demonstrated using annealing qubits with coherence times around 1 $\mu$s~\cite{Weber2017}. 

\subsubsection{3D integration: coherence, connectivity, and tailored annealing. }

\begin{figure} %
\centering
\includegraphics[width=1.0\textwidth]{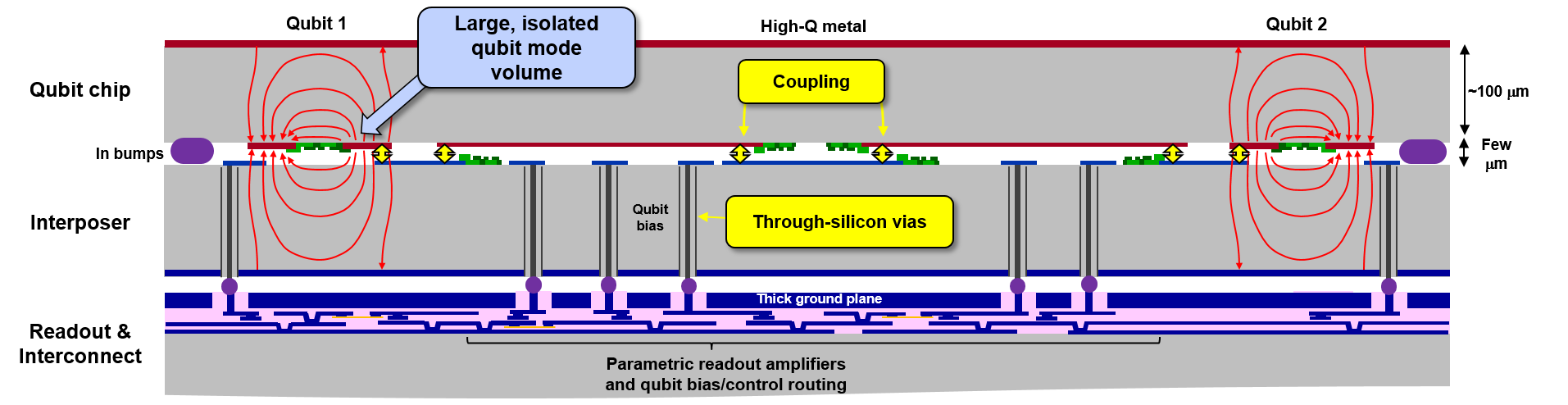}
\caption{An approach to 3D integration that avoids the pitfalls of monolithic integration. Each chip in the stack -- qubit chip, interposer, and readout \& interconnect chip -- is fabricated independently. The chips are then joined using indium bump-bonding. The approach enables high-coherence qubits in conjunction with addressibility.} 
\label{fig:3D-integration}
\end{figure}

Current approaches to building quantum annealers generally use monolithic, multi-layer fabrication processes. The advantage of this approach is that it supports high levels of complexity, high connectivity, and the annealing protocol can be implemented on chip using superconducting control electronics, such as is done for the D-Wave 2000Q. However, this type of multi-layer fabrication exacerbates low coherence times due to the many fabrication steps required and the use of interwiring layer amorphous dielectrics. The result is an abundance of materials and fabrication-induced defects that sap coherence from the qubits. 

An alternative approach was proposed in Ref.~\cite{Rosenberg2017}, and is called the ``three stack'' (see Fig.~\ref{fig:3D-integration}). In this approach, there are three layers with unique functionalities, and each is fabricated separately. The qubit chip is fabricated in a high-coherence process similar to those used for gate-model qubits. The readout and interconnect layer is a multi-layer process similar to conventional annealing processes; it can support dense interconnects, high connectivity, and active devices such as parametric amplifiers~\cite{Macklin2015} and SFQ electronics~\cite{Tolpygo2014}. To isolate this layer from the qubit chip, an interposer layer is introduced. The interposer is made of low-loss, intrinsic silicon and features superconducting through-silicon vias to bring signals from the readout and interconnect layer up to the qubit plane. By keeping the readout and interconnect layer far from the qubit chip, the electric and magnetic fields associated with the qubits are constrained to the low-loss qubit and interposer layers. The layers are then joined together using standard indium bump-bonding techniques. 

With the three-stack or similar approaches, one can controllably study the impact of high coherence, high connectivity, and non-stoquastic coupling on the quantum annealing and its potential to achieve quantum advantage. Each layer can be fabricated independently and then combined in a manner that is itself extensible. This approach also supports both on-chip and off-chip control electronics, to enable annealing protocols that are tailored to each qubit and coupler individually. Once the key to achieving quantum advantage is understood, one can then focus on building larger machines using the most appropriate approach(es).

\subsection{Trapped ions} 

While the mainstream platform for quantum annealing is superconducting qubits, with impressive progress in recent years, there do exist alternative systems. 
In particular, since atomic levels do not suffer from fabrication errors, quantum simulation platforms based on neutral or charged atoms represent promising alternatives in order to implement quantum annealers for practical applications or as highly-controlled test beds for fundamental questions. 
Examples include Rydberg atoms (see next section) and trapped ions, which we discuss in this section. 

Charged ions \cite{kielpinski2002architecture} can be trapped by electromagnetic confining potentials in high vacuum, e.g., in linear Paul traps or Penning traps. The ions can then be cooled almost to the absolute ground state of their motional degrees of freedom, through which they arrange in regular crystal patterns due to their mutual Coulomb repulsion.   
The constituent qubits are encoded as two electronic hyperfine states, on which single-qubit rotations can be realized with extremely high fidelity by laser light or microwaves, depending on the chosen hyperfine states. These single-qubit terms can have arbitrary spatial dependence, thanks to single-ion addressing \cite{Jurcevic2014,Smith2016,Maier2019}. 
Due to the strength of the Coulomb repulsion, the distance between individual ions is typically on the order of a few micrometers, over which direct interactions between the hyperfine qubits are negligible. To introduce effective qubit--qubit interactions, one can couple the qubits off-resonantly via laser or microwave radiation to the collective phononic vibrations of the ion crystal \cite{cirac1995quantum,Sorensen1999,Mintert2001}. Eliminating the phonons in second-order perturbation theory, the resulting interaction yields an Ising Hamiltonian with long-ranged coupling terms $J_{ij}$ \cite{Britton2012, Islam2012, Richerme2013, Jurcevic2014, Nevado2014, Trautmann2018}.

Thus, including single-qubit rotations, the natural Hamiltonian for trapped-ion systems is 
\begin{equation}
    \label{eq:ionHamiltonian}
    H=\sum_{i\neq j} J_{ij}(t) \sigma_i^z\sigma_j^z + \sum_i \sum_{\beta=x,y,z} B_i^\beta(t) \sigma_i^\beta\,. 
\end{equation}
Time dependence of the Hamiltonian parameters is controlled simply by ramping laser intensities. This Hamiltonian thus provides all the ingredients for a basic annealing protocol. 

As one limiting factor, the spatial dependence of the interactions $J_{ij}$ is determined by the laser parameters, dimensionality of the crystal, and the phonon modes. 
If a single frequency $\mu$ is used to generate the interactions, they are typically given by \cite{Richerme2013}
\begin{equation}
\label{eq:Jij}
J_{ij}\propto \Omega_i \Omega_j \sum_q \frac{\eta_i^q \eta_j^q}{\mu^2-\omega_q^2}  .  
\end{equation}
Here, $\Omega_i$ is the laser Rabi frequency at ion $i$, $\omega_q$ is the frequency of the phonon mode $q$, and $\eta_i^q$ is determined by the amplitude of the vibrational mode $q$ at ion $i$.

\begin{figure} %
\centering
\includegraphics[width=1.0\textwidth]{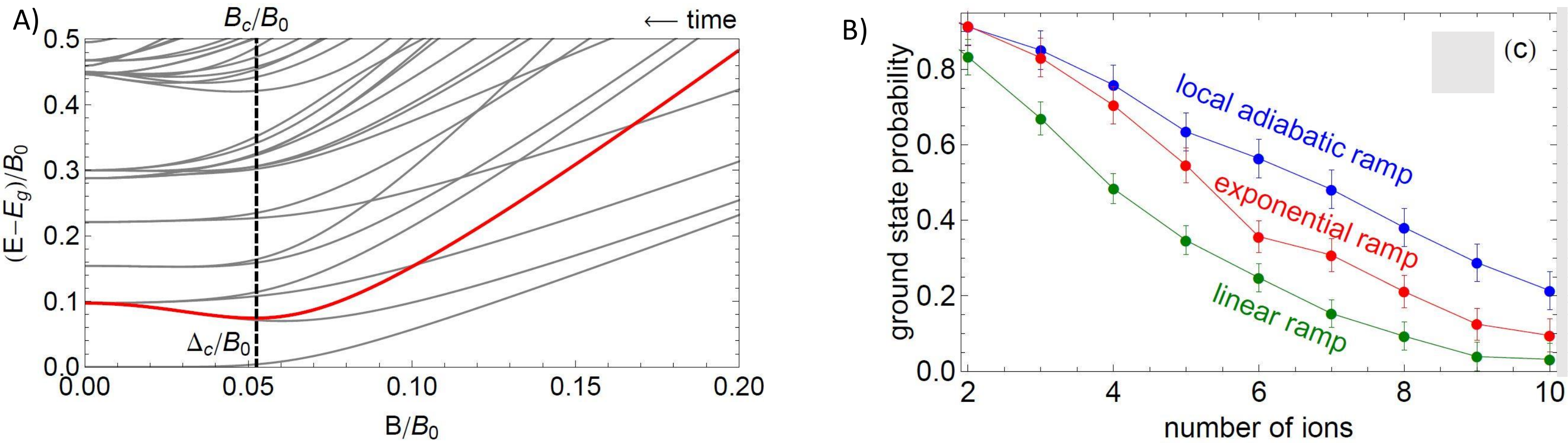}
\caption{
Adiabatic preparation of a spin-model ground in trapped ions, as reported in Ref.~\cite{Richerme2013}. The model is described by the Hamiltonian $H=\sum_{i\neq j} J_{ij} \sigma_i^z\sigma_j^z + B(t)\sum_i  \sigma_i^x$. 
(A) Energy levels $E$ relative to the ground state $E_g$, computed for $N=6$. The thick red line denotes the first excited level that gets coupled to the ground state and which defines the minimum gap $\Delta_c$. The ramp begins at $B(t=0)\equiv B_0\gg \max(J_{ij})$, which also sets the energy scale. 
(B) For experimentally accessible ramp times, the overlap of the final state with the ground state rapidly decreases with system size. This effect can be mitigated by optimizing the ramp, for example, by being slower close to the minimum gap (denoted local adiabatic). Taken from Ref.~\cite{Richerme2013}. 
} 
\label{fig:adiabatic_ions}
\end{figure}

In typical implementations, the frequency that generates the spin-spin interactions is tuned outside of the phonon band. For small systems, 
\begin{equation}
J_{ij}\propto \frac{1}{|i-j|^\alpha}    
\end{equation}
can then be approximated by a tunable powerlaw in the range $\alpha\in [0,3]$ \cite{Trautmann2018}, while for larger systems it is better represented by a combination of an exponential and a dipolar decrease with distance \cite{Nevado2014}. 
Following a groundbreaking adiabatic protocol involving only two spins \cite{Friedenauer2008}, quantum annealing with such long-range interactions and homogeneous fields has been realized with up to 18 qubits \cite{Islam2012,Richerme2013} (although phrased in the context of quantum simulation and adiabatic state preparation of spin-model ground states). Though the trade-off between annealing speed and decoherence limited the overlap to the ground state at the end of the sweep, see Fig.~\ref{fig:adiabatic_ions}, these experiments opened the road towards adiabatic protocols using trapped ions. 

To proceed further on the road towards applications in quantum annealing, one requires besides engineering improvements (see below), the ability to freely program a desired optimization problem. 
In Ref.~\cite{Hauke2015}, it was shown that Hamiltonian Eq.~(\ref{eq:ionHamiltonian}) is sufficient to encode Coulomb glass problems as well as a variety of NP-complete models in a completely programmable manner, including the knapsack problem, number partitioning, and instances of the max-cut problem. 
To do this, one requires in addition to the programmable local rotations $B_i^\beta$ only the ability to locally adjust the intensity $\Omega_i$ in Eq.~(\ref{eq:Jij}). Both are experimentally feasible with existing single-site addressing-techniques using acousto-optic modulators or digital-mirror devices \cite{Smith2016,Maier2019}.  
An alternative approach to obtain benchmark models for quantum annealing is by generating deterministic instances of disordered spin-glass problems. This can be realized by tuning the radiation frequency $\mu$ into the phonon band \cite{Grass2016}, which generate interactions with high randomness and thus glassy behavior. Although the models resulting in this way may not be freely programmable, they can nevertheless provide valuable tests for quantum annealing protocols. Finally, if a number of frequencies that scales with the number of qubits is used to induce the qubit--qubit interactions, arbitrary interaction patterns $J_{ij}$ can be programmed \cite{Korenblit2012}. 

Regarding trap design and qubit coherence, there has been strong progress since the seminal experiments of Ref.~\cite{Friedenauer2008,Islam2012,Richerme2013}. Quantum simulations in linear Paul traps are now performed with tens of ions \cite{Zhang2017,Friis2018}. Cryogenic traps will allow experimentalists to scale to 100 ions or more \cite{Pagano2018}. Numbers of this order are already being achieved in Penning traps \cite{Britton2012}, where adiabatic protocols in the Dicke model (an infinite-range spin model coupled to a bosonic mode) have recently been attempted for $\sim 70$ ions \cite{Safavi-Naini2018}. Segmented traps \cite{Zippilli2014} and microfabricated surface traps \cite{Seidelin2006,Schmied2009b,Hakelberg2018,Boldin2018} also open new possibilities for the design of interaction patterns. All the while, trapped-ion experiments are characterized by extremely large degrees of coherence. Life-times of the qubit levels are on the order of several seconds. Current limiting factors are fluctuations of magnetic fields and laser intensities. Often, these limit coherence times during the dynamics to several milliseconds, a time scale that is comparable to the one set by typical  interaction strengths on the order of several kHz \cite{Zhang2017,Safavi-Naini2018}. By use of decoherence-free subspaces, these times can be increased to tens of milliseconds \cite{Jurcevic2014,Maier2019}. 

As an interesting feature in this context, trapped ions allow for testing the effect of decoherence on the annealing protocol in a controlled manner. For example, by letting the local fields fluctuate in a temporally random way, dephasing noise can be simulated \cite{Soare2014,Hauke2015}. Recently, it has been demonstrated that practically arbitrary noise spectral functions can be realized in this way, which has been used to study the effect of noise on quantum transport in a toy model for a photosynthetic complex \cite{Maier2019}. It may be interesting to extend such analyses to true "quantum baths", e.g., by resonant coupling to phonon modes as has been realized in a small toy model in Ref.~\cite{Gorman2018}. Thus, the trapped-ion platform may provide a valuable test bed to benchmark quantum annealing subject to designed forms of decoherence. 

\subsection{Neutral atoms} 
\label{sec:rydberg}

Another ideal platform for quantum simulation in and out of equilibrium is Rydberg atoms, which offer large coherence times and individual control of qubits \cite{SaffmanReview}. Using so-called dressing schemes \cite{jaksch2000fast}, Rydberg atoms can implement Ising-spin models \cite{Schauss2015,Maller2015,Jau2016,Labuhn2016} with nearest neighbor interactions. In the context of adiabatic quantum computing, a natural question is which models can be directly implemented in current neutral atom platforms, given that interactions are induced by laser-fields, which are challenging to tune individually. 

In Ref.~\cite{glaetzle2017coherent}, the authors propose a Rydberg implementation for coherent quantum annealing based on the LHZ scheme \cite{lechner2015quantum}, see Eq.~(\ref{eq:LHZ_Hamiltonian}). In this square lattice scheme, interactions are problem independent [$C_n$ in Eq.~(\ref{eq:LHZ_Hamiltonian})] and can thus be applied using a few lasers globally among all atoms. Local fields, which encode the optimization problem [$J_i$ in Eq.~(\ref{eq:LHZ_Hamiltonian})], can be applied with digital mirror devices with single site resolution. The main challenge in implementing LHZ is the realization of the required 4-body ZZZZ interactions. In Ref.~\cite{glaetzle2017coherent}, this issue is solved by a decomposition of the 4-body interactions into 2-body interactions with an overhead of one ancilla per plaquette. These ancilla atoms are required to be of a different species (e.g., by combining Rb and Cs atoms), which poses an experimental challenge. 

A direct implementation of the maximum independent set problem in neutral atoms was recently introduced in Ref.~\cite{pichler2018quantum}. The maximum independent set of a graph is defined as the maximum set of vertices where no pair of vertices in each subset is connected by an edge. An experimental simplification can be achieved when restricting to unit disk graphs, which still leaves the problem NP-complete. These are planar graphs where vertices are placed on a two-dimensional grid and edges between vertices with a pair-wise distance smaller than a certain threshold. This problem can be directly mapped to Rydberg atoms using the so-called blockade mechanism, by which there exists a blockade radius around any atom that is excited to a Rydberg state within which no other atoms can be excited. Thus, the proposal serves as a direct test of quantum optimization without overhead from embedding. 

As pointed out in Sec.~\ref{sec:optimization problems}, a main challenge in quantum annealing is the infinite range interactions typical for hard optimization problems. In Ref.~\cite{torggler2018quantum}, the authors propose a direct implementation of an infinite range system with atoms in a cavity. The encoded problem is a variation of the $N$-Queens problem. The $N$-Queens problem is: given an $N \times N$ chess-board, place $N$ queens such that no pieces attack each other (queens attack along rows, columns, and diagonals). A particular variation, the blocked-diagonals variation, is NP-complete and cannot be solved with current classical algorithms for $N>21$. The problem can be mapped to an annealing protocol with $N$ $N$-level systems. Each $N$-level system represents a column of the chess-board and each level a row. For example, if the second level of the third $N$-level system is occupied it means that a queen is placed on the second row of the third column on the chess board. The $N$-level systems are realized with atoms in one-dimensional optical lattices and the non-attacking condition is implemented with strong repulsive interactions along diagonals and rows. These interactions are mediated by the optical cavity and infinite range. The proposal is an attractive test bed for direct implementations of optimization problems. In particular, for this problem classical algorithms fail already for small numbers of atoms that will be available in near term. The proposal of Ref.~\cite{torggler2018quantum} is based on coherence times limited to the order of $100$ms and interaction strengths of $Jt=50$, which is the same order of magnitude as already demonstrated in experiment \cite{landig2016quantum}. 

\section{Summary\label{sec:discussion}}

With the advent of intermediate scale quantum devices, we are currently witnessing the start of the second quantum revolution, where quantum mechanics will be used as a tool for computations tasks. With its relative simplicity and reduced demands on resources, quantum annealing may serve as one of the first crucial stepping stones towards full-fledged quantum computing. In this perspectives article, we have summarized possible routes towards solving the open questions in the field, including the question of possible quantum speedup from novel protocols, new encoding strategies that enable high coherent qubit platforms for quantum annealing, and novel hardware architectures that push the limits of coherence and control. The field is still open and evolving rapidly, and we hope this article serves as a guideline for future research paving the way for quantum technology as a computational tool.  

\section*{Acknowledgements}
P.H.~acknowledges support by the DFG Collaborative Research Centre SFB 1225 (ISOQUANT) and the ERC Starting Grant StrEnQTh.  
The work of H.G.K. and W.D.O.~is supported in part by the Office of the Director
of National Intelligence (ODNI), Intelligence Advanced Research Projects
Activity (IARPA), via MIT Lincoln Laboratory Air Force Contract
No.~FA8721-05-C-0002. 
W.L. acknowledges funding by the Austrian Science Fund (FWF) through a START grant under Project No. Y1067-N27 and the SFB BeyondC Project No.~F7108-N38, the Hauser-Raspe foundation, and the European Union's Horizon 2020 research and innovation program under grant agreement No.~817482 PasQuanS.
The work of H.N.~is based upon work partially supported by the Office of the Director of National Intelligence (ODNI), Intelligence Advanced
Research Projects Activity (IARPA), via the U.S. Army Research Office
contract W911NF-17-C-0050. The views and conclusions contained herein are
those of the authors and should not be interpreted as necessarily
representing the official policies or endorsements, either expressed or
implied, of ODNI, IARPA, or the U.S.~Government.  The U.S.~Government is
authorized to reproduce and distribute reprints for Governmental purpose
notwithstanding any copyright annotation thereon. 
\\

\providecommand{\newblock}{}

\end{document}